\pdfoutput=1 
\documentclass[longauth]{aa}  
\usepackage{graphicx}
\graphicspath{ {./Figures/} }
\usepackage{txfonts}
\usepackage{amssymb}
\usepackage{amsmath}
\usepackage[english]{babel}

\usepackage{xcolor}
\usepackage{nicefrac}
\usepackage{stfloats}
\usepackage{multirow}
\usepackage{booktabs}
\usepackage{siunitx}
\usepackage{verbatim}
\usepackage{ragged2e}       

\usepackage{hyperref}
\hypersetup{colorlinks = true,
            linkcolor = blue,
            urlcolor  = blue,
            citecolor = blue,
            anchorcolor = blue}

\makeatletter
\renewcommand*\aa@pageof{, page \thepage{} of \pageref*{LastPage}}
\makeatother

\newcommand{\Msun}{M_\odot }
\newcommand{\Mjup}{\mathrm{M_J} }

\newcommand{\Ha}{$H\alpha$ }
\newcommand{\Qphi}{$ Q_\phi $ }
\newcommand{\secname}{Sec.}
\newcommand{\eqname}{Eq.}
\newcolumntype{C}{>{$}c<{$}}    

\begin{document} 

    \title{ Disk Evolution Study Through Imaging of Nearby Young Stars (DESTINYS):  
    HD\,34700\,A unveils an inner ring 
    \thanks{
    Based on observations collected at the European Southern Observatory under ESO programme(s) 1104.C-0415(A), 0108.C-0583(A) and 0108.C-0583(B).}
    }
    
    \titlerunning{HD\,34700\,A unveils an inner ring }
    \authorrunning{ G. Columba et al.}

    \author{G.~Columba \inst{1,2}
          \and
          E.~Rigliaco \inst{2}
          \and
          R.~Gratton \inst{2}
          \and
          D.~Mesa \inst{2}
          \and
          V.~D'Orazi \inst{2, 3}
          \and
          C.~Ginski \inst{4}
          \and
          N.~Engler \inst{5}
          \and 
          J.~P.~Williams \inst{6}
           \and 
          J.~Bae \inst{7}
           \and 
          M.~Benisty \inst{8,9}
           \and 
          T.~Birnstiel \inst{10,11}
           \and 
          P.~Delorme \inst{8}
           \and
          C.~Dominik \inst{12}
           \and 
          S.~Facchini \inst{13}
           \and 
          F.~Menard \inst{8}
           \and 
          P.~Pinilla \inst{14}
           \and 
          C.~Rab \inst{10,15}
           \and 
          Á.~Ribas \inst{16}
          \and 
          V.~Squicciarini \inst{2,17}
           \and 
          R.\,G.~van Holstein \inst{18}
           \and 
          A.~Zurlo \inst{19,20,21}
    }

    \institute{ 
        Department of Physics and Astronomy "Galileo Galilei" - University of Padova, Vicolo dell’Osservatorio 3, 35122 Padova, Italy \\
        \email{gabriele.columba@phd.unipd.it}
        \and
        INAF – Osservatorio Astronomico di Padova, Vicolo dell’Osservatorio 5, 35122 Padova, Italy 
        \and
        Dept. of Physics - University of Rome Tor Vergata, Via della Ricerca Scientifica 1, 00133 Rome, Italy 
        \and 
        School of Natural Sciences, Center for Astronomy, University of Galway, Galway, H91 CF50, Ireland 
        \and
        ETH Zürich - Institute for Particle Physics and Astrophysics, 
        Wolfgang-Pauli Str. 27, 8093 Zürich, Switzerland
        \and
        Institute for Astronomy, University of Hawaii at Manoa, Honolulu, HI 96822, USA
        \and
        Department of Astronomy, University of Florida, Gainesville, FL 32611, USA
        \and
        Univ. Grenoble Alpes, CNRS, IPAG, F-38000 Grenoble, France
        \and
        Université Côte d’Azur, Observatoire de la Côte d’Azur, CNRS, Laboratoire Lagrange, France
        \and
        University Observatory, Faculty of Physics, Ludwig-Maximilians-Universität München, Scheinerstr. 1, 81679 Munich, Germany
        \and
        Exzellenzcluster ORIGINS, Boltzmannstr. 2, D-85748 Garching, Germany
        \and
        Anton Pannekoek Institute for Astronomy, University of Amsterdam, Science Park 904, 1098XH Amsterdam, The Netherlands
        \and
        Dipartimento di Fisica, Universit\`a degli Studi di Milano, Via Celoria, 16, Milano, I-20133, Italy
        \and
        Mullard Space Science Laboratory, University College London, Holmbury St Mary, Dorking, UK
        \and
        Max-Planck-Institut für extraterrestrische Physik, Giessenbachstrasse 1, 85748 Garching, German
        \and
        Institute of Astronomy, University of Cambridge, Madingley Road, Cambridge, CB3 0HA, UK
        \and
        LESIA, Observatoire de Paris, Université PSL, CNRS, 5 Place Jules Janssen, 92190 Meudon, France
        \and
        European Southern Observatory, Alonso de C\'{o}rdova 3107, Casilla 19001, Vitacura, Santiago, Chile
        \and
        Instituto de Estudios Astrof\'isicos, Facultad de Ingenier\'ia y Ciencias, Universidad Diego Portales, Av. Ej\'ercito Libertador 441, Santiago, Chile
        \and
        Escuela de Ingeniería Industrial, Facultad de Ingeniería y Ciencias, Universidad Diego Portales, Av. Ejercito 441, Santiago, Chile
        \and
        Millennium Nucleus on Young Exoplanets and their Moons (YEMS)
    }


 
    \abstract
    { The study of protoplanetary disks is fundamental to understand their evolution and interaction with the surrounding environment, and to constrain planet formation mechanisms. }
    { We aim at characterising the young binary system HD\,34700\,A, which shows a wealth of structures.}
    { Taking advantage of the high-contrast imaging instruments SPHERE at the VLT, LMIRCam at the LBT, and of ALMA observations, we analyse this system at multiple wavelengths. We study the rings and spiral arms morphology and the scattering properties of the dust. We discuss the possible causes of all the observed features. } 
    { We detect for the first time, in the \Ha band, a ring extending from \SI{\sim 65}{\astronomicalunit} to \SI{\sim 120}{\astronomicalunit}, inside the ring already known from recent studies. These two have different physical and geometrical properties. Based on the scattering properties, the outer ring may consist of grains of typical size $a_\mathrm{out} \geq \SI{ 4}{\um} $, while the inner ring of smaller grains ($a_\mathrm{in} \leq $ \SI{0.4}{\um}). Two extended logarithmic spiral arms stem from opposite sides of the disk. The outer ring appears as a spiral arm itself, with a variable radial distance from the centre and extended substructures.
    ALMA data confirm the presence of a millimetric dust substructure centred just outside the outer ring, and detect misaligned gas rotation patterns for HD\,34700 A and B. 
    }
    {  
    The complexity of HD\,34700\,A, revealed by the variety of observed features, suggests the existence of one or more disk-shaping physical mechanisms.
    Possible scenarios, compatible with our findings, involve the presence inside the disk of a yet undetected planet of several Jupiter masses and the system interaction with the surroundings by means of gas cloudlet capture or flybys. Further observations with JWST/MIRI or ALMA (gas kinematics) could shed more light on these.}

    \keywords{ protoplanetary disks -- intermediate-mass stars -- planet formation -- multiple systems -- direct imaging }

    \maketitle
%

\section{Introduction}\label{sec:intro}

    Recent years have seen an exponential growth in the number of discovered exoplanets and we are now beyond the five thousands milestone\footnote{ See the NASA Exoplanet Archive: \tiny \url{ https://exoplanetarchive.ipac.caltech.edu/index.html} }. 
    However, detecting substellar companions within infant systems disks has proven to be a much harder task. In fact, the presence of dust and gas left from stellar formation can hide the already faint light from planets and brown dwarfs \citep{Szulagyi2019, Sanchis+2020, Szulagyi&Garufi2021}. Therefore, it is not surprising that to date only in one system giant planets were clearly detected within its transition disk, PDS\,70 \citep{Keppler+2018:PDS70, Mesa+2019:PDS70, Haffert+2019:PDS70}. 
    Recently, evidence for strongly embedded protoplanets was presented (\citealt{Boccaletti+2020} and \citealt{Currie+2022} for AB\,Aur, \citealt{Hammond+2023} for HD\,169142), demonstrating the efforts required to detect these very young objects at near-infrared and optical wavelengths.
    The extensive study of protoplanetary disks is crucial to understand the earliest phases of planet formation, drawing an evolutionary path from the raw materials of the disk to the final features of a fully formed system. 
    
    In this context, we have studied in depth the young system HD\,34700\,A (HIP\,24855), which shows complex substructures. The stellar and disk parameters available in the literature are summarised in \tablename~\ref{tab:target}.
    This system started to receive attention since the early 2000s: \cite{Arellano&Giridhar2003:HD34700} first determined that HD\,34700\,A was a double-lined spectroscopic binary of G0IVe type \citep{Mora+2001:vega-like}. 
    \cite{Torres2004:HD34700} confirmed the binarity of the system and calculated for it an orbit with eccentricity $ e = 0.25 $ and a period of $ \sim 23.5 $ days. They found that the two stars have almost equal mass and estimated their orbit to be seen at an inclination of $ \sim \ang{39}$. They also detected youth indicators: a weak \Ha emission, a strong \ion{Li}{I} absorption, a strong X-ray emission and high rotational velocities, all supporting the hypothesis of a pre-main sequence system, of few tens of Myr. 
    \cite{Sterzik+2005:HD34700} discovered that HD\,34700 is actually a multiple system of at least 4 stars: the central binary (A) and two low-mass companions of the M stellar type, projected at distances of \ang{;;5.23} (B) and \ang{;;9.25} (C). They also found the radial velocities of B and C to be consistent with the A component, and detected in all their spectra youth indicators such as \ion{Li}{I} and \Ha emission.
    
    More recently,  HD\,34700\,A disk was resolved for the first time by \cite{Monnier+2019:HD34700}, through polarised $H$-band observations with the Gemini Planet Imager (GPI; \citealt{Macintosh+2014:GPI}), showing its wide ring, a cavity, several spiral arms and an inner faint arc. They carried out a thorough analysis of HD\,34700\,A disk, comparing the data to radiative transfer models, to reproduce the observed ring-like structure. They obtained for the ring an average radius of \SI{175}{\astronomicalunit} at an inclination of \ang{42} to the line of sight (LoS).
    They noticed that the ring projected semi-major axis and the peak polarisation axis are misaligned, corresponding to position angles (PAs) respectively of \ang{69} and \ang{86} East of North (EoN). Moreover, \cite{Monnier+2019:HD34700} provided updated estimates of the system parameters, based on the new accurate parallax value: they confirmed HD\,34700\,A to be a \SI{\sim 5}{Myr}-old binary star with a total mass around \SI{4}{\Msun} and a transition disk. 
    \cite{Uyama+2020:HD34700} imaged again the outer ring of HD\,34700\,A with Subaru/CHARIS in JHK bands and focused on the characterisation of its morphology, spiral arms included, employing angular differential imaging \citep[ADI;][]{Marois+2006:ADI} reduction techniques. They found the ring to have a PA of \ang{74.5} and an inclination of \ang{40.9}, by ellipse-fitting, and its centre to be offset from the star of \SI{52.7}{mas} towards \ang{110.8}. While no substellar companion was found, they assessed upper limits for any such object of around \SI{5}{\Mjup} outside the ring and \SI{12}{\Mjup} in the cavity.
    At longer wavelengths, \cite{Benac+2020:HD34700} presented Submillimeter Array (SMA) observations of HD\,34700 A and B. Their observations of the \element[][12]{CO} J=2--1 transition line detected a counterclockwise Keplerian rotation of A's disk, while the \SI{1.3}{\mm} continuum resolved a prominent azimuthally asymmetric emission, they suggest could be evidence for a dust trap. This substructure appears to peak right outside the scattered light ring, at around \SI{155}{\astronomicalunit} from the central star, based on their deprojection. 
    They resolved continuum emission around the B component too, caused by its own circumstellar disk.
    Thanks to the improved parallaxes obtained as part of the \emph{Gaia Data Release 3} \citep{GaiaCollab2022:DR3}, the latest estimate place HD\,34700\,A at distance of \SI{350.5}{pc}.
    
\begin{figure}
        \centering
        \includegraphics[ width=\hsize]{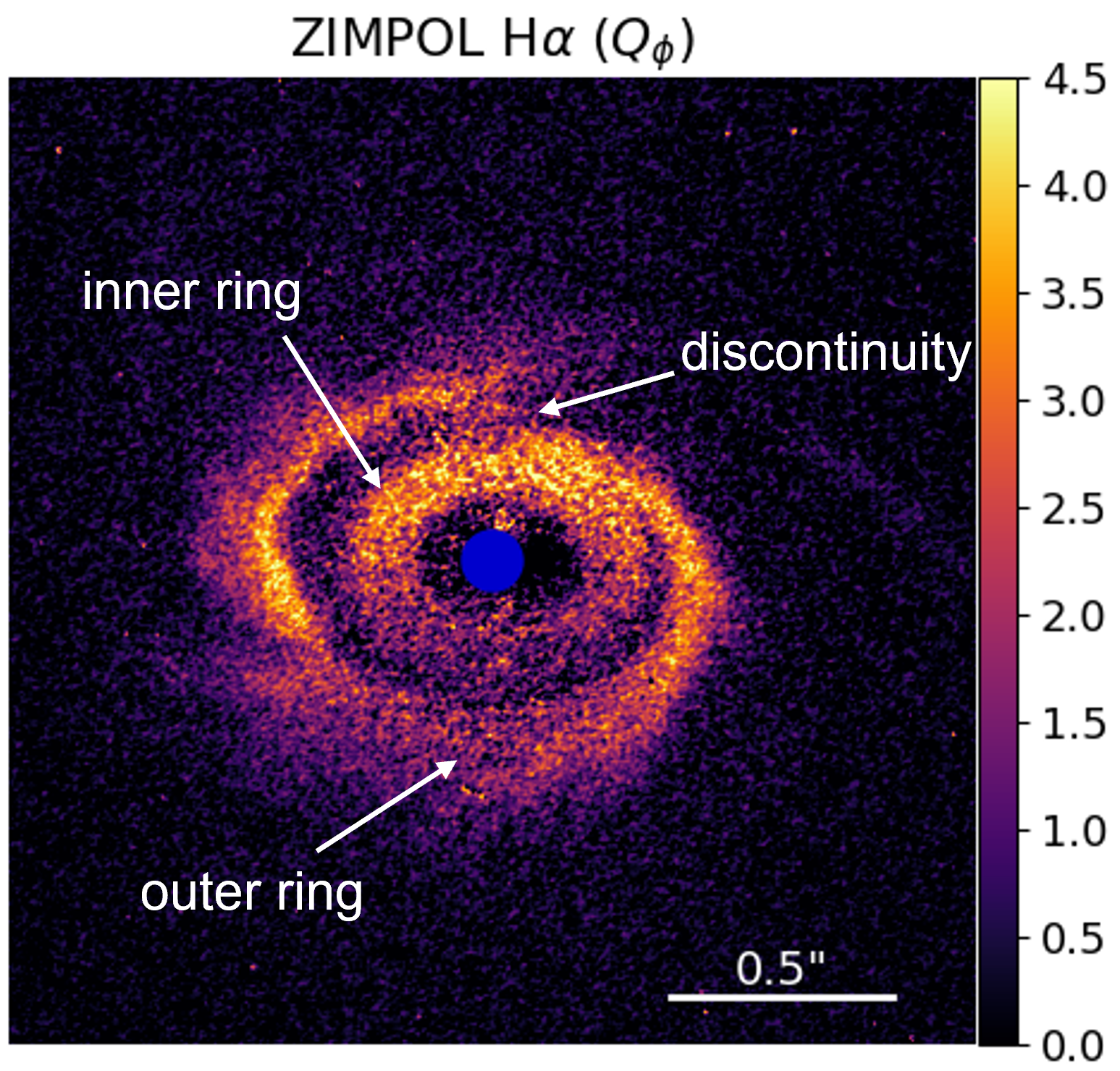}
        \caption{ ZIMPOL \Qphi image of HD\,34700\,A with annotations. The field of view is 2.16$^{\prime\prime}\times2.16^{\prime\prime}$. The blue circle represents the coronagraph. North is up and East is left. The colour bar is linear and has arbitrary units. }
        \label{fig:Qphi_ZIMPOL}
    \end{figure}

    In this work we present new observations of HD\,34700\,A and analyse its various features, discussing the implications of a new ring within the large cavity, not fully resolved in previous observations of \cite{Monnier+2019:HD34700} and \cite{Uyama+2020:HD34700}. 
    \figurename~\ref{fig:Qphi_ZIMPOL} illustrates the main features that are analysed and discussed: the inner ring (detected for the first time), the outer ring, and its discontinuity. 
    We describe the new observational data that we have obtained in \secname~\ref{sec:data} and dive into the analysis of the disk components in \secname~\ref{sec:analysis}, highlighting what insight we can gain from the new data. Then, in \secname~\ref{sec:discuss} we discuss the interpretation of our results, also in comparison to previous studies on this object. Finally, we summarise and conclude in \secname~\ref{sec:conclude}.

    \begin{table}[hbt]
        \caption{Summary of known parameters of HD\,34700\,A. The "bin" subscript indicates the A binary parameters.}             
        \label{tab:target}      
        \centering
        \begin{tabular}{l c c}
            \toprule     
            Parameter   &   Value   &   Reference   \\ 
            \midrule                  
            R.A. (J2000)    &   05:19:41.41       &    1   \\
            DEC. (J2000)    &   +05:38:42.80       &    1   \\
            Distance        &   $ 350.5\ (\pm 2.5) $ \si{pc}   &   1   \\
            $ P_{\mathrm{bin}} $    &   23.4877 days     &    2   \\
             $ e_\mathrm{bin} $               &           0.25             &    2   \\
            $ i_\mathrm{bin} $  &       \ang{39}    &   2       \\
            $ T_{\mathrm{eff}} $ &  5900 K + 5800 K      &    2   \\
            $ q \equiv M_1 / M_2 $ &  $ 0.987 \pm 0.014 $ & 2 \\
            $ v \sin{i} $ (Aa, Ab)   &   $ 26.3,\ 21.7\ (\pm 0.6) $ \si{\km\per\s}  &    3   \\
            Spectral type        &   G0 IVe + G0 IVe     &    4   \\
            Age             &   \SI{ \sim 5}{\mega yr}      &    5   \\
            $ M_\mathrm{bin} $        &   \SI{ 4}{M_\sun}    &    5   \\
            $ M_\mathrm{dust} $        &   1.2 $ (\pm 0.2)  \times 10^{-4} $ M$_\sun$    &    5   \\
            $ i_\mathrm{outring} $        &   \ang{42}    &    5   \\
            \bottomrule 
        \end{tabular}
        \tablebib{ 
            (1) \cite{GaiaCollab2021:GeDR3}; 
            (2) \cite{Torres2004:HD34700};
            (3) \cite{Sterzik+2005:HD34700};
            (4) \cite{Mora+2001:vega-like};
            (5) \cite{Monnier+2019:HD34700}; 
        }
    \end{table}

\section{Observations and data reduction}
    \label{sec:data}

    HD\,34700\,A system has been the target of several observations in recent years. In the following we present all the new observations acquired, and the data reduction process. A short summary of all the observations and the relative setups is reported in \tablename~\ref{tab:obslog}.

    \begin{table*}
        \centering
        \caption{Observation setup and observing conditions.}
        \label{tab:obslog}
        \begin{tabular}{@{}lcccccccc@{}}
            \toprule
            Date  		& Instrument & Mode & Filter	    & Coronagraph	   & DIT [s] & \# Frames & Seeing [arcsec] & $\tau_0$ [ms]\\
            
            \midrule
            27-10-2019	& SPHERE & IRDIS-DPI &    BB\_H		& YJH\_ALC	& 64.0      & 68	&	0.71	&	4.6 \\
            14-11-2021	& SPHERE & IRDIS-DBI &    K1/K2		& YJH\_ALC	& 16.0	  & 66		&	0.76	&	5.9	\\
            14-11-2021	& SPHERE & IFS &    YJH		& YJH\_ALC	& 32.0	  & 15		&	0.74	&	4.2	\\
            26-11-2021	& SPHERE & IRDIS-DBI &    K1/K2		& YJH\_ALC	& 32.0	  & 104		&	0.76	&	7.7	\\
            26-11-2021	& SPHERE & IFS &    YJH		& YJH\_ALC	& 32.0	  & 14		&	0.75	&	7.3	\\
            08-12-2021	& SPHERE  & ZIMPOL (slow pol) &    B\_\Ha		& CLC	&  100	  & 20		&	0.46	&	10.7	\\
            14-02-2022 & LBT & LMIRCam & L$^{\prime}$ & none & 0.81	  &  14284		&	0.92	&	--	\\
            21-08-2022 & ALMA & -- & Band 6 & none & -- & -- & -- & -- \\
            \bottomrule
        \end{tabular}
    \end{table*}

    \subsection{SPHERE observations}
        
        HD\,34700\,A was observed with SPHERE, the Spectro-Polarimetric High-contrast Exoplanet REsearch facility \citep{Beuzit+2019:SPHERE}, mounted at the UT3's Nasmyth focus of ESO Very Large Telescope (VLT), with all its scientific systems: IRDIS (the Infra-Red Dual-beam Imager and Spectrograph, \citealt{Dohlen+2008:irdis, Vigan+2010:irdis}), IFS (the Integral Field Spectrograph, \citealt{Claudi+2008:IFS, Mesa+2015:IFS}) and ZIMPOL (the Zurich Imaging Polarimeter, \citealt{Schimd+2018:zimpol}).  
        The SPHERE data were acquired within the program \emph{Disk Evolution Study Through Imaging of Nearby Young Stars} \citep[DESTINYS;][]{Ginski+2020:destinys, Ginski+2021:SUaur}, aimed at studying young planet-forming disks in scattered light. 
        To determine the star position behind the coronagraph in the science frames, additional "centre frames" were taken for all instruments, using the satellite spots obtained in the deformable mirror waffle mode.

        \subsubsection{IRDIS polarised data}
        
            SPHERE/IRDIS data was acquired on 27 October 2019 in broadband $H$ (BB\_H, $\lambda_c $=\SI{1.625}{\um}, $\Delta\lambda$=\SI{0.290}{\um}, and pixel scale \SI{12.251}{mas \ pix^{-1}}, \citealt{Maire+2016:sphere}). 
            The observations were carried out in dual beam polarimetric imaging mode \citep[DPI;][]{deBoer+2020:irdisDPI, vanHolstein+2020:irdisDPI}, with a coronagraphic mask of radius \SI{92.5}{mas} \citep{Carbillet+2011, Guerri+2011}, centred on the photocentre of the binary star. The observations consisted of 14 polarimetric cycles, each containing four exposures taken at half-wave plate (HWP) switch angles 0$^{\circ}$, 45$^{\circ}$, 22.5$^{\circ}$, and 67.5$^{\circ}$. The individual frame exposure time was set to 64\,s, resulting in a total integration time of 74 min. 
            The data were reduced with the IRDAP tool\footnote{ The default settings were adopted for this reduction. See the software documentation at: \url{https://irdap.readthedocs.io/en/latest/}} \citep{vanHolstein+2020:irdisDPI}, which applied polarimetric differential imaging (PDI, \citealt{Kuhn+2001}), producing the images corresponding to the Stokes parameters $Q$ and $U$, \Qphi and $ U_\phi $, shown all together for completeness in \figurename~\ref{fig:IRDIS_quad}. The \Qphi image, particularly useful to our goals, is shown alongside the other data in \figurename~\ref{fig:all_new_wave} (third panel). 
            The azimuthal Stokes parameters \Qphi and $U_\phi$ were obtained according to the definitions of \citet{deBoer+2020:irdisDPI}:
            \begin{equation}
            \begin{array}{@{}l@{}}
                Q_\phi = - Q\cos( 2\phi ) - U\sin( 2\phi ) \\
                U_\phi = Q\sin( 2\phi ) - U\cos( 2\phi ) \\
            \end{array}
            \end{equation} 
            where $\phi$ is the EoN polar angle in the coordinate system centred on the star. 
    
            Polarimetry is well suited for the study of circumstellar disks, allowing us to isolate their signal from the excessive stellar brightness. The central stars mostly emit non-polarised light, which can be easily distinguished from the polarised light scattered by the dust in the disk. 
            The observing strategy was pupil tracking \citep{vanHolstein2017:irdis}, which allows the rotation of the field of view (FoV) in time with respect to the detector.
            \figurename~\ref{fig:Qphi_r2} shows the $H$-band \Qphi $r^2$-scaled image of HD\,34700\,A, that accounts for the $r^{-2}$ dependency of the stellar flux, after considering the inclination of the disk.

        \subsubsection{IRDIS+IFS data}
            \label{subsec:irdifs_obs}
        
            New data in \emph{IRDIFS\_EXT} mode were acquired on 14 and 26 November 2021.
            This mode uses simultaneously IRDIS and IFS, to join the advantages of simultaneous imaging and spectroscopy. With our setup IFS covered the wavelength range 0.95-\SI{1.65}{\um} at a spectral resolution of R$\sim$30 (FoV 1.77 arcsec$^2$), and IRDIS covered the $K1$ and $K2$ bands (dual-band imaging mode, or DBI, see \citealp{Vigan+2010:irdis}, FoV of $\sim$10 arcsec$^2$) at \SI{2.09}{\um} and \SI{2.22}{\um}, respectively. We used the N\_ALC\_YJH\_S apodized Lyot coronagraph \citep{Boccaletti+2008} with a field mask having a radius of \SI{92.5}{mas}.  
            The observing strategy for this set of data was star-hopping in pupil-stabilised mode \citep[see e.g.][]{Wahhaj+2021:starhop}. The total field rotation angle was \ang{16.6} and \ang{42.7} for the two epochs. 
            Star-hopping consists of the acquisition of a template Point Spread Function (PSF) from a chosen nearby reference star, in between the exposure of the scientific target. Subtracting the reference PSF from the science target has two main benefits. Firstly, at small separation one actually gains sensitivity thanks to the lack of signal self-suppression (as there is in ADI). Secondly, ADI distorts disk morphology due to self-subtraction and can sometimes lead to false positive detections \citep[see, e.g.,][]{Rameau+2017}.
            
            We applied two techniques for the suppression of the stellar speckle field in the IRDIFS\_EXT data. The IRDIS \textit{K}-band imaging data utilised the iterative reference star differential imaging (iRDI) approach briefly outlined in \cite{Ginski+2021:SUaur}. In this case the reference star data observed during star hopping is used as the library of PSF standards. To fit the reference star data to the science data we use the KLIP (\citealt{Soummer+2012:PCA}) principal component analysis (PCA) code. Since the disk is bright relative to the stellar speckle field, we need to prevent overfitting of the reference star data. This is done iteratively by feeding the positive disk signal above a specified threshold back into the routine and subtracting it from the science data before the reference star images are fitted. For the reduction of HD\,34700 data we used a threshold of 100 counts, which includes bright disk signal but excludes remaining speckle noise. In each step we removed 15 KLIP-modes and we ran a total of 100 iterations. We note that this technique is basically identical to the iterative ADI approach presented in \cite{Stapper&Ginski2022:iADI}, with the exception that a reference star is used for speckle suppression instead of the science data itself. 
            
\begin{figure*}     
            \centering
            \includegraphics[width=0.33\hsize]{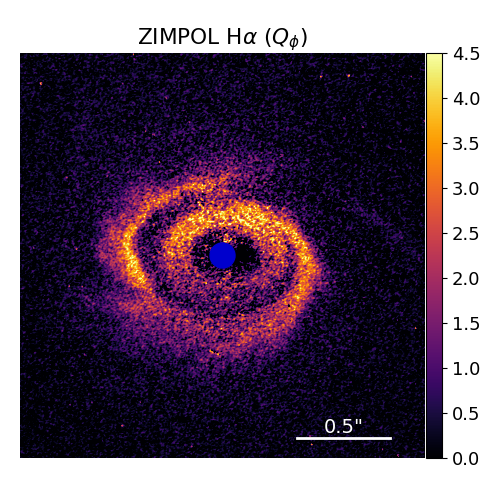}
            \includegraphics[trim=0 8 0 8, clip, width=0.34\hsize]{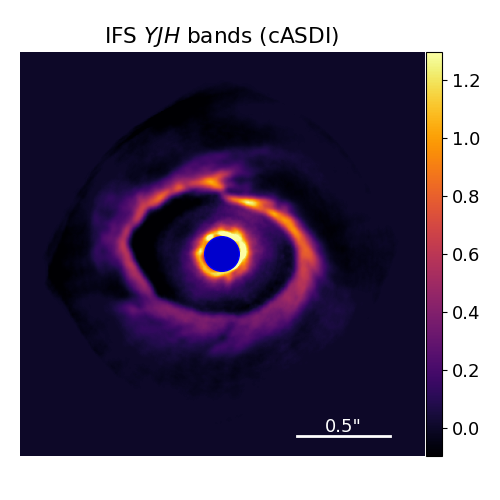} 
            \includegraphics[width=0.32\hsize]{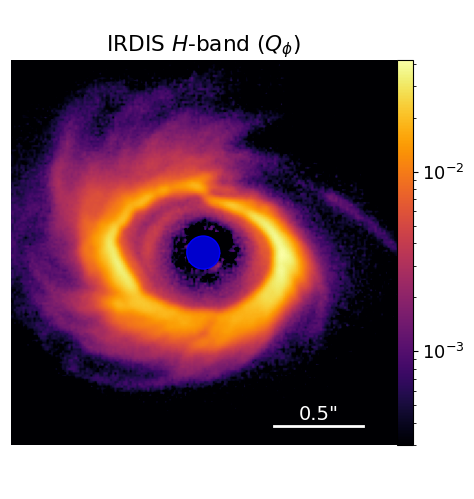}
            \\
            \includegraphics[trim=0 10 2 8, clip, width=0.33\hsize]{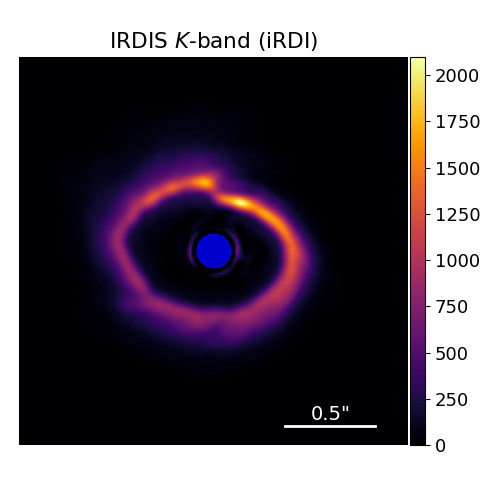}            
            \includegraphics[trim=0 10 4 8, clip, width=0.33\hsize]{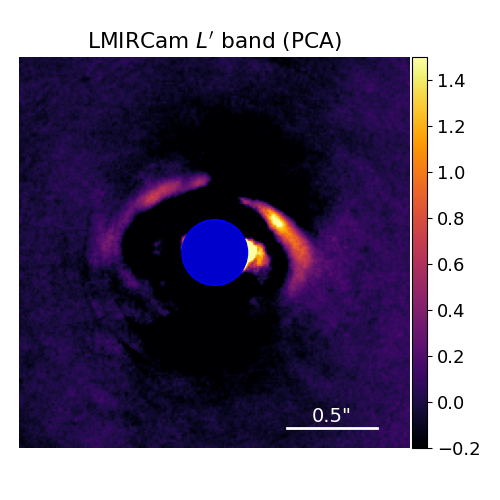}
            \includegraphics[trim=30 20 0 8, clip, width=0.324\hsize]{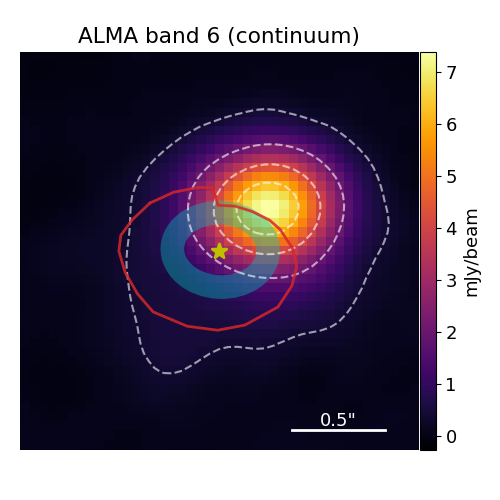}
            
            \caption{ Images of HD\,34700\,A showing the observational data used in this paper, in order of increasing wavelength. North is up and East is to the left. The FoV shown in all panels is 2.16$^{\prime\prime}\times2.16^{\prime\prime}$. The blue circles represent the coronagraphs, except for LMIRCam observations in L$^{\prime}$ band (without coronagraph), where the blue area masks all the spurious signals within $ \lambda/D $, and ALMA data. 
            In the ALMA continuum image the red line represents the outer ring, while the blue ellipse the inner ring; the continuum beam size is $0\farcs41\times0\farcs28$ at a PA of \ang{-64}. The dashed contours denote levels of [0.4, 2, 4, 6] mJy/beam. The colour bars have arbitrary units except for ALMA; they all have linear scale, except for IRDIS $H$-band, which is logarithmic.}
            \label{fig:all_new_wave}
        \end{figure*}

            For the IFS data we use a combination of classical angular and spectral differential imaging (cASDI). The idea of cASDI is to overcome the self-subtraction inherent to classical ADI and SDI individually by combining the spectral and temporal dimensions of the IFS data. We compute the median PSF to subtract from all images simultaneously in the temporal and the spectral dimension. We first re-scale all wavelength channels so that the speckle fields coincide. We then flatten this array across all parallactic angles and compute the median. By including both the parallactic angle rotation as well as the wavelength scaling, we increase the diversity within the data set such that less disk signal remains in the median image used for subtraction compared to classical ADI or SDI.
            The total intensity \textit{K}-band iRDI and \textit{YJH}-bands cASDI images are shown in the fourth and second panel of \figurename~\ref{fig:all_new_wave}.

        \subsubsection{ZIMPOL data}

            ZIMPOL observations were acquired with excellent sky conditions (see \tablename~\ref{tab:obslog}) on 7-8 December 2021, in the 
            B\_Ha ($\lambda_c$=\SI{655.6}{nm}) and Cnt\_Ha ($\lambda_c$=\SI{644.9}{nm}) \Ha filters. 
            ZIMPOL achieves diffraction-limited performance in these bands, on a FoV of $ 3.5'' \times 3.5''$. 
            These observations were performed in the slow polarimetry P1 mode of the instrument. In five polarimetric $QU$ cycles the consecutive measurements of the Stokes parameters $Q^+$, $Q^-$, $U^+$, and $U^-$ were conducted with different HWP switch angles of $0^\circ$, $45^\circ$, $22.5^\circ$, and $67.5^\circ$, respectively. 
            At the beginning of the science observation the flux measurement was performed with the star offset from the coronagraphic mask, using the neutral density filter ND\_1.0 and DIT = \SI{10}{\s}. 
    
            The ZIMPOL data were reduced with the ETH data reduction pipeline (described in, e.g., \citealt{Schimd+2018:zimpol, Hunziker+2020:zimpol}). Pre-processing and calibration of raw frames included subtraction of bias frames, flat-fielding, and correction for the modulation and demodulation efficiency. The instrumental polarisation was corrected through forced normalisation of the fluxes in the frames of two opposite polarisation states. All frames were centred at the position provided by the centre frames. The final format of the reduced $Q$ and $U$ images is $1024\times 1024$ pixels with a pixel size of approximately $3.6 \times 3.6$\,mas on sky. 
            The final \Qphi image of the ZIMPOL \Ha data, shown in \figurename~\ref{fig:Qphi_ZIMPOL}, is the mean of the independent \Qphi frames from the two ZIMPOL cameras, for the B\_Ha filter.

    \subsection{LBT observations}
    
        New data were also acquired with LBT/LMIRCam (the Large Binocular Telescope Mid-InfraRed interferometric Camera, \citealt{Skrutskie+2010:LBT}) on 14 February 2022 as part of the program IT-2021B-AO-3 (P.I. Gratton). The observations log is reported in \tablename~\ref{tab:obslog}. 
        LMIRCam's field of view of $ 11.0'' \times 11.0'' $, similar to IRDIS, was wide enough to include both the system's A and B components in the frames. The observation was performed in pupil stabilised mode, with the $ L' $ band filter (effective wavelength \SI{3.7}{\um}, \citealt{Skrutskie+2010:LBT}), with a final scientific exposure time of \SI{3.22}{\hour} (\SI{0.81}{\s} each frame), obtaining a final rotation of the FoV of \ang{75.4} that is enough to effectively perform the ADI reduction.
        The high thermal background was removed from the data by subtracting a median of the two nodding positions adopted in the observing strategy. Then, the disk was detected applying the PCA (see e.g. \citealt{Soummer+2012:PCA}) algorithm, using a custom-made code appositely prepared to reduce these data. To this aim, only one principal component was used to create the image to be subtracted from the science images. The final image is shown in the second to last panel of \figurename~\ref{fig:all_new_wave}.

    \subsection{ALMA observations}

        The ALMA observations were taken in ALMA Band 6 (240\,GHz, 1.25\,mm) on 21 August 2022 as part of the DESTINYS snapshot program 2021.1.01705.S (PI C. Ginski). The integration time on source was only 1.5 minutes in moderate weather conditions with precipitable water vapor levels of 0.35\,mm. The baseline lengths ranged from 15\,m to 1200\,m. The data were calibrated using the Common Astronomy Software Applications package  (CASA, \citealt{2007ASPC..376..127M}), version 6.2.1, to first remove atmospheric and instrumental effects using the quasar calibrators. We then self-calibrated the data to increase the signal-to-noise ratio.

        The correlator was configured with broad wavelength coverage for best continuum sensitivity but included one spectral window centred on the 230.538\,GHz lines of \element[][12]{CO} $J=2-1$ in the lower sideband.
        The continuum image was produced using the \emph{tclean} task on all spectral channels other than the CO window. The image shown in the last panel of \figurename~\ref{fig:all_new_wave} was produced with Briggs robust weighting parameter $-0.5$ and corrected for primary beam attenuation. 
        The beam size is $0\farcs41\times0\farcs28$ at a PA of \ang{-64} and the rms is 0.13\,mJy\,beam$^{-1}$.
        The \element[][12]{CO} spectral window was imaged channel by channel (0.65\,km\,s$^{-1}$) with a Briggs robust equal to $+0.5$ and a resulting beam size $0\farcs52\times0\farcs48$ at a PA of \ang{-61} with an rms of 4.3\,mJy\,beam$^{-1}$.

\section{Analysis}   
    \label{sec:analysis}

    In the following, we analyse the disk structures seen in the new observations, compare them to the previous ones and highlight the new insights on the system's morphology.
    In \figurename~\ref{fig:all_new_wave} we illustrate all the new multi-band data presented in this study.
    From a comparison of these images, it becomes clear that we can divide the disk around HD\,34700\,A in two main features: the outer disk, already detected and analysed in recent years, which is very bright in all wavelengths; and an inner disk, whose first detection was possible thanks to the \Ha-band polarimetry.

    \subsection{Rings fit}
        \label{subsec:ringfit}

\begin{figure}      
            \centering
            \includegraphics[trim=4 0 0 0, clip, width=\hsize]{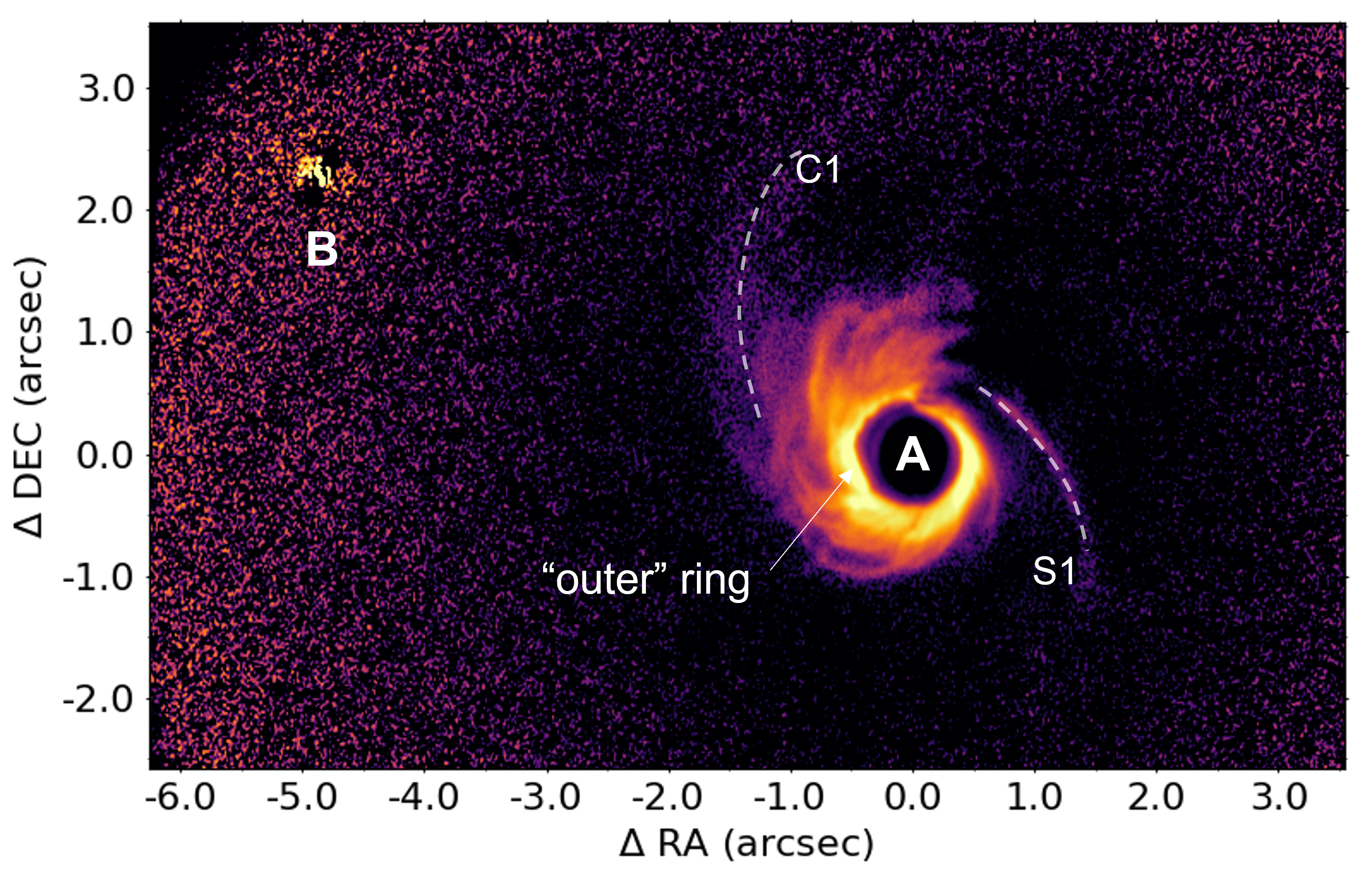}
            \caption{ Face-on deprojection of IRDIS H-band \Qphi data, corrected by $r^2$ light dilution, and with special logarithmic colour scale to highlight the faint outer spirals. The two opposite arms at East and West (C1 and S1) stand out from the disk. HD\,34700\,B component is also visible in the FoV, in the upper-left corner. The inner ring plane ($ i \sim \ang{40} $, see \tablename~\ref{tab:inring}) was considered for deprojecting to face-on.
            }
            \label{fig:Qphi_r2}
        \end{figure}

        \textbf{ Outer ring.} 
        The scattered light from the outer dust ring is clearly visible in all datasets of \figurename~\ref{fig:all_new_wave}, except for the ALMA continuum.
        Multiple spiral arms structures outside a large cavity, characterised by a strong discontinuity near \ang{0} EoN, were already detected in $J$- and $H$-band images, obtained with GPI and SCExAO by \cite{Monnier+2019:HD34700} and \cite{Uyama+2020:HD34700}, respectively. 
        IRDIS images (\figurename~\ref{fig:Qphi_r2} and \figurename~\ref{fig:IRDIS_quad}) confirm the presence of these structures. 
        Our ALMA data (last panel of \figurename~\ref{fig:all_new_wave}) confirmed that the ring discontinuity is near the asymmetric mm-dust structure first discovered by \cite{Benac+2020:HD34700}. 
        The nearest side of the disk to the observer is expected to be the Northern side, based on \cite{Monnier+2019:HD34700} and confirmed by our observations. 
        In this work we do not provide new estimates for the outer ring PA and inclination, as it was already done by previous works; instead, we attempt different interpretations, based on the new perspectives offered by the inner ring discovery. Nevertheless, we do provide an analysis of its scattering phase function (see \secname~\ref{subsec:phasefunc}) and of the spiral structures tied to the outer ring (see \secname~\ref{subsec:spirals}).

        \begin{figure}      
            \centering
            \includegraphics[trim=0 30 0 50, clip, width=\linewidth]{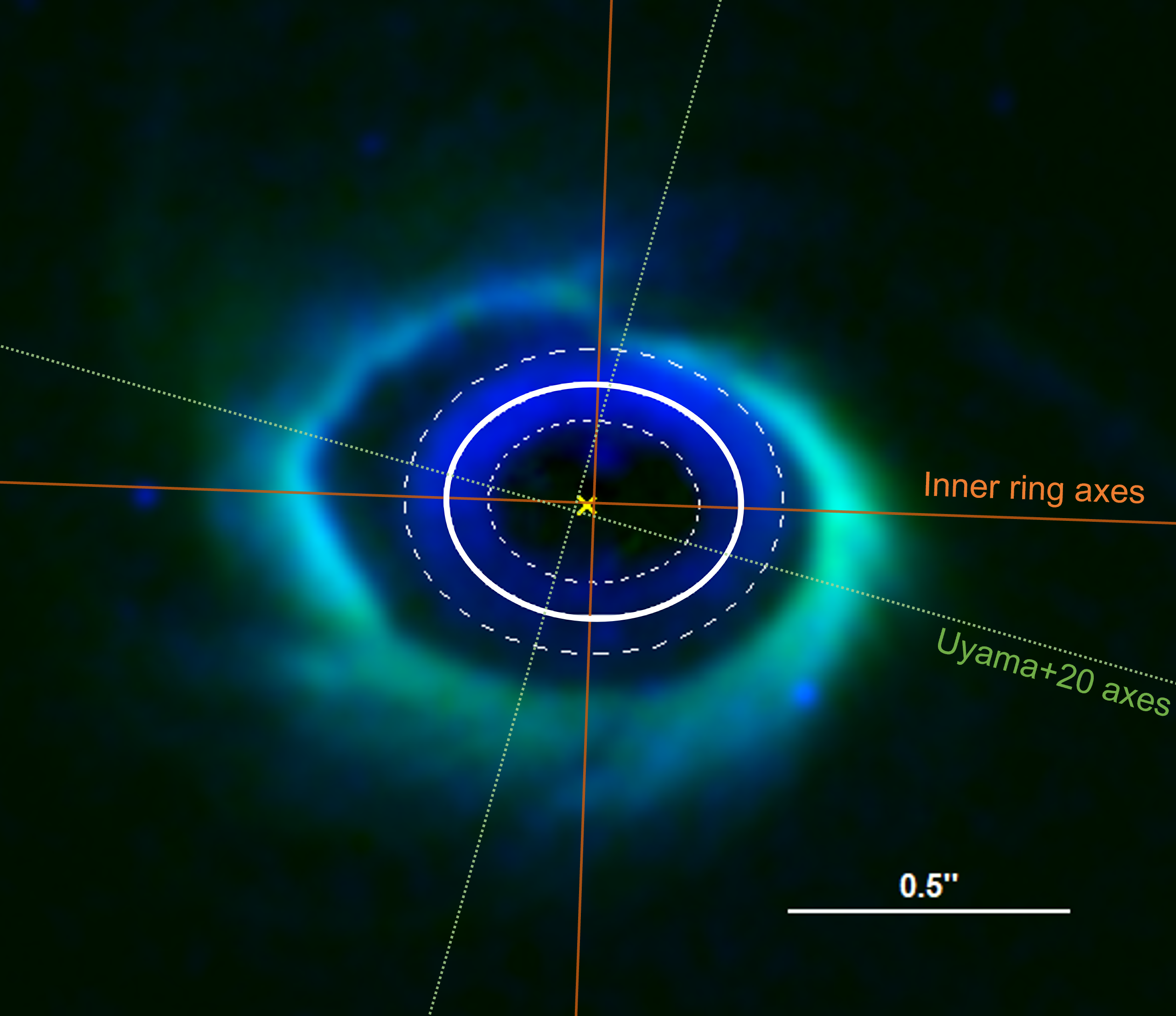}
            \caption{ Overlay of the ellipse axes (\emph{orange}) and the inner ring (\emph{white}) as fitted in this work. The yellow cross marks the position of the central binary star. For reference we included the ellipse axes obtained by \cite{Uyama+2020:HD34700} based on the outer ring alone (\emph{dotted green}). 
            The underlying image is a two-band \Qphi composite of the IRDIS H-band (\emph{green layer}) and the ZIMPOL \Ha (\emph{blue layer}). }
            \label{fig:ellipsefit}
        \end{figure}

        \textbf{ Inner ring.}
        Polarimetric observation from ZIMPOL in the \Ha band (\figurename~\ref{fig:Qphi_ZIMPOL}), revealed the clear presence of another ring structure located inside the outer ring cavity. 
        This inner ring extends radially from $\sim${65}\,au to $\sim$120\,au (or $ 186-\SI{335}{mas} $) from its centre. It appears to have an offset with respect to the nominal position of the central star (more on the offsets in \secname~\ref{subsec:offsets}). 
        It is intriguing to notice the presence of a faint "arc" inside the outer ring, towards N-E, already detected by \citet{Monnier+2019:HD34700}, that we found again in the IRDIS \Qphi image (see \figurename~\ref{fig:IRDIS_quad}). Based on the new ZIMPOL data, we can link this arc structure to the inner ring as they are found to overlap well and trace the same elliptical shape. 
        Furthermore, the inner ring seems to partially merge with the outer ring structure on the N-W side, especially near the step-like discontinuity around \ang{0}. Depending on the disk 3D geometry, however, this intersection may be more a perspective effect than a physical contact.

        We characterised the geometry of the inner ring by assuming a circular dust ring and fitting two separate ellipses to its internal and external boundaries (see \figurename~\ref{fig:ellipsefit}).
        We selected the ring boundaries by subtracting, from a slightly smoothed (14-17.6\,mas) version of the original image, a strongly smoothed image (42.2-45.7\,mas). The result of this subtraction yields positive values for the pixels of the bright ring core and negative values on the surrounding background. Then, we chose as the boundaries the set of coordinates corresponding to a sign change in this subtraction image.
        This method proved to be the most suitable for this system, since the brightness fluctuates over scales of many pixels, creating fuzzy boundaries. Also, we excluded from the fit the North-West quadrant points, as the two rings overlap significantly there. 
        The selected boundary points were successively fed to the "least-squares ellipse fitting" algorithm developed by \cite{Hammel&Molina2020}\footnote{\tiny \url{https://github.com/bdhammel/least-squares-ellipse-fitting/tree/v2.0.0} } in \texttt{Python}. 
        The whole procedure was repeated by varying the angle of the inner ring mask and the smoothing parameters. 
        Finally, we computed the best-fit values as the mean of the two independent boundaries, taking the median for each parameter and having the standard deviations as the uncertainty. 
        These single-ellipse parameters are the ones employed in our subsequent deprojections.
        We have summarised the resulting parameters and errors in \tablename~\ref{tab:inring}. 
        We obtained an inner ring with a semi-major axis PA of \ang{87 \pm 1} from the North direction, seen at an inclination of \ang{\sim 40 \pm 2}. 
        The PA value differs by 15-\ang{20} from the outer ring values obtained by \cite{Monnier+2019:HD34700} and \cite{Uyama+2020:HD34700}. However, \cite{Monnier+2019:HD34700} already noticed that the peak polarisation axis of the outer structure was found at \ang{86}, which is very close to the value we have obtained from the inner ring. The LoS inclination we found is also compatible with previous estimates.

        \begin{table}[bt]
            \caption{Geometrical parameters of HD\,34700\,A inner ring. }             
            \label{tab:inring}      
            \centering
            \setlength{\tabcolsep}{12pt}
            \begin{tabular}{l c c}
                \toprule     
                Parameter   &   Value  \\ 
                \midrule  
                Inner radius   & $65\pm0.6$~(au)     \\
        	Outer radius   & $117\pm0.2$~(au)    \\
        	Mean width     & $52\pm0.7$~(au)     \\ 
        	Mean PA        & $87.1\pm1.22$~(°)  \\
                LoS inclination & $39.7\pm2.0$~(°) \\
                Centre offset\tablefootmark{a} & $9.8\pm5.7$~(mas) \\
                Centre PA\tablefootmark{a} & $-52.3 \pm15.4$~(°) \\
        
                \bottomrule 
            \end{tabular}
            \tablefoot{
                \tablefoottext{a}{with respect to the nominal position of HD\,34700\,A.} \\
                The reader can find in the Introduction the parameters fit by \citet{Uyama+2020:HD34700} for the outer ring.
            }
        \end{table}

        The two rings appear different. The external one has spiral arms of various shapes and sizes, and it shows a sharp discontinuity. On the contrary, the internal ring has a sharper boundary and appears as a highly regular elliptical annulus. 
        The differences between these two rings, however, are not limited to their shape: while the inner ring is prominent in the \Ha band, it is almost undetected in the $H$-band. 
        There is only a faint arc in the N-E direction that can be seen in the IRDIS $H$-band \Qphi (bottom left panel of \figurename~\ref{fig:IRDIS_quad}), overlapping with the inner ring edge. This faint feature was already detected by \cite{Monnier+2019:HD34700}, who noticed that the cavity inside the outer ring was not completely devoid of dust scattering light. 
        
        The fact that the inner ring was fully revealed at shorter wavelengths implies an underlying segregation between the dust grain sizes, highlighting a deeper physical distinction. We discuss the possible formation mechanisms of such structures in \secname~\ref{subsec:dust_segreg}.

    \subsection{Geometrical offsets}
        \label{subsec:offsets}

        In the following, we analyse the geometrical offsets observed for the centres of the inner and the outer ring. 

        \subsubsection{Inner ring offset}
        
        HD\,34700\,A shows offsets of the inner/outer rings centres with respect to the central star.  
        The centre of the elliptical annulus traced by the inner ring, as described in \secname~\ref{subsec:ringfit}, shows a shift from the star of around \SI{3}{\astronomicalunit} towards the West and \SI{2}{\astronomicalunit} towards the North. Precisely, we find this geometrical centre to be found at a PA of \ang{-52 \pm 15} and distance of \SI{9.8 \pm 5.7}{mas} (\SI{3.4 \pm 2.0}{\astronomicalunit}) from HD\,34700\,A nominal position, as reported also in \tablename~\ref{tab:inring}. The uncertainties provided originate from the $1\sigma$ errors on the inner ring geometrical fit, plus the half-pixel star-centring error, causing a relatively large but realistic uncertainty. 
        The direction of these offsets was rather unexpected. Offsets are common for flaring disks along the minor axis, due to the disk height and image projection \citep[see, e.g.,][]{Ginski+2016, deBoer+2016}. The offsets we retrieved for the inner ring are opposite of what would be expected based on that: surface flaring should have shifted the centre towards the South, whereas we find a vertical shift towards the North. There is also a significant component along the major axis direction. These features suggest that other mechanisms could be responsible for the observed offsets: the inner ring is likely to be rather flat and eccentric. 
        We notice in this context that the PA of the inner ring centre has a value very close to the PA of the supposed mm-dust trap. From the ALMA continuum, in fact, we find a $PA_\mathrm{mm} \sim \ang{-48 \pm 11}  $ measuring the angle of the 95\% peak intensity seen from the central star. This $PA_\mathrm{mm}$ is only \ang{\sim 4} away from the inner ring centre PA (\ang{-52}). We speculate on a correlation between these two angles, but a reliable answer would require detailed theoretical simulations beyond the scope of this work.

        \subsubsection{Outer ring offset}
        
        For the outer ring, assuming that the observed structure is a projection from a circular ring, \cite{Uyama+2020:HD34700} reported a shift of the outer ellipse centre of around \SI{-17.3}{\astronomicalunit} in the West direction and \SI{-6.6}{\astronomicalunit} in the North direction from the central star. This is exactly opposite to what we find from the inner ring (but in the direction which can be justified by height projection). We did not fit the outer ring as a regular ellipse, thus we do not provide updated estimates on its offsets and parameters.
        Given the compatible position angles and inclinations between inner and outer ring, the different parts of the outer ring may not be equidistant from the central star (see the spiral fit in \secname~\ref{subsec:spirals}). Also, because of possible azimuthal and radial variations in scale height, the light scattered by the outer dust may not belong to a unique geometrical plane. Thus, the circular approximation might not be particularly suited for the outer ring.

        \subsubsection{Unseen stellar companions}
            \label{subsec:psfsearch}
            
        Having ruled out the inner disk flaring as a plausible cause of the observed offsets, we investigated both the potential presence of an additional stellar companion and a non-null eccentricity of the inner ring. 
        We have searched for other stellar companions in the vicinity of the central binary, massive enough to bring the barycentre away from the photocentre. 
        We used the calibration frames saved during the observing runs to reveal potential close point-source companions hidden behind the coronagraph. 
        We processed the "no-coro" calibration frames from IFS, performing a simple ADI-like reduction by subtracting the median PSF of the central binary and adding the de-rotated frames, as in \citet{Bonavita+2022:shine}.
        The maximum field rotation (42.7$^{\circ}$, see also \secname~\ref{subsec:irdifs_obs}), obtained for the 26-11-2021 calibration frames, allows the revelation of potential companions beyond $ 1 \text{rad} / \ang{42.7} \times \lambda / D = 42\,$mas to be free from self-subtraction. The post-processed IFS data provides a J magnitude contrast of $\sim 7\,$mag at \SI{50}{mas}, which corresponds to a limit in mass of \SI{50}{\Mjup}, using the BHAC15 models \citep{Baraffe2015A&A}. The full contrast limits plot and the processed SNR map are shown in the Appendix, in \figurename~\ref{fig:nocoro_lims} and \figurename~\ref{fig:nocoro_map} respectively. 
        We do not detect any reliable point-source in the processed image, hence, we can definitely exclude the presence of unseen stellar companions. Indeed, everything more massive than 30-\SI{50}{\Mjup} behind the coronagraph would have been detected with this reduction. 
        
        We can reinforce and expand this argument with other considerations. 
        Firstly, the SED of HD\,34700\,A shows an infrared excess, which must be caused by an innermost circumbinary dust ring, not visible due to its proximity to the central star and heated up to high temperatures. \cite{SeokLi2015:HD34700} fit results suggested a disk starting around \SI{0.3}{\astronomicalunit} with a $ T \SI{\sim 1500}{\kelvin} $ (they had an imprecise system parallax, however). Thus, any stellar companion close to the central binary would irremediably perturb the inner disk edge, causing issues to the infrared excess. Secondly, if a stellar companion were further out of the central binary, we would have detected it easily, being outside of the coronagraph. 
        There are also astrometric arguments against this stellar companion. The Gaia RUWE\footnote{"Renormalized Unit Weight Error", it indicates the quality of the astrometric fit. A RUWE > 1.4 could indicate that the source is non-single (or otherwise problematic) and it is commonly used as a possible hint of unresolved companions.} parameter for HD\,34700\,A of 1.04 is fully consistent with a single object (and the central binary is tight enough to have no effect on the RUWE parameter). Moreover, the proper motion anomaly (PMa) measured by \cite{Kervella+2022:pma} is low and compatible with substellar masses, thus not significant for a star. In fact, this excludes the presence of a stellar companion with a mass above \SI{0.2}{\Msun} in the separation range \SIrange[]{3}{30}{\astronomicalunit}, the latter being approximately the size of the coronagraph used in our SPHERE observations. 
        All these points suggest that there is no third stellar companion in the range \SIrange{2}{30}{\astronomicalunit} that could have shifted the barycentre of this system.

        \subsubsection{Ring eccentricity}
        
        Lastly, we assume the inner ring is eccentric and estimate its eccentricity from its offset. The binary star would be the centre of mass of the ring, positioned in one of the two foci, which means that the offset between the ellipse centre and the star is a function of the eccentricity and can be used backwards to infer $ e $. 
        To estimate the eccentricity, we approximated the semi-major axis as the same obtained from the geometrical fit described above. For the centre of the ellipse we considered the displacements in both directions, after deprojecting to the inner ring plane. 
        We find for the focal distance $ c = 3.2 \pm 1.0 \,$pix and for the semi-major axis $ a = \SI{74.0 \pm 0.6}{pix} $, corresponding to a modest ring eccentricity of $ e = c/ a = 0.043 \pm 0.014 $. 
        Taking into account this eccentricity in the projection effects along the LoS, it would result in a smaller inclination angle $ i $ than for the circular scenario. The difference, however, is below one degree less than what found in \tablename~\ref{tab:inring} and within the error bars. 
               
        Hydrodynamical simulations of protoplanetary disks demonstrate that objects heavier than $\sim$\SI{3}{\Mjup} lead to significantly eccentric cavities in the dust \citep[see, e.g.,][]{Kley&Dirksen2006:ppds, Pinilla+2012b:dust, Zhang+2018:dsharp}.
        Specifically, \citet{Kley&Dirksen2006:ppds} provide disk eccentricity values as a function of radius, for different companion masses (or, better, mass ratios). They report a peak eccentricity of $ \sim 0.05 $ for $ q = 0.001 $ (\SI{4}{\Mjup} in HD\,34700\,A) for orbits both inside and outside the planet. Heavier companions excite much higher eccentricities in the disk outside their orbit (up to $0.22$), but again around $ \gtrsim 0.05 $ for inner orbits. Comparing these theoretical results with our eccentricity estimate, we find a good agreement. A \SI{\sim 4}{\Mjup} companion may be present either inside or outside HD\,34700\,A inner disk. Heavier planets are unlikely to reside inside the inner ring cavity, as they would excite much higher eccentricities; they could be present anyway in the gap between the inner and the outer rings. 
        In addition, \cite{Zhang+2018:dsharp} show that disk eccentricity is enhanced by low viscosity and small aspect ratios. 
        We know that the aspect ratio of HD\,34700\,A cannot be too high, since the offsets along the projected minor axis are limited. A low-viscosity coefficient, moreover, appears to be preferred among protoplanetary disks, as reported by the DSHARP collaboration \citep{Dullemond+2018:dsharp}, but also \citet{Pinte+2022, Rosotti2023}. 
        
        In conclusion, the presence of a giant planet of several Jupiter masses, which cannot be ruled out by our observations, would naturally explain the offset issue by the excitation of disk eccentricity. 
        Further discussion on a potential substellar object and its impact on dust segregation is reported in \secname~\ref{subsec:dust_segreg}.

    \subsection{Spiral arms fit}
        \label{subsec:spirals}

        \begin{figure*}
            \centering      
            \includegraphics[ width=0.9\hsize]{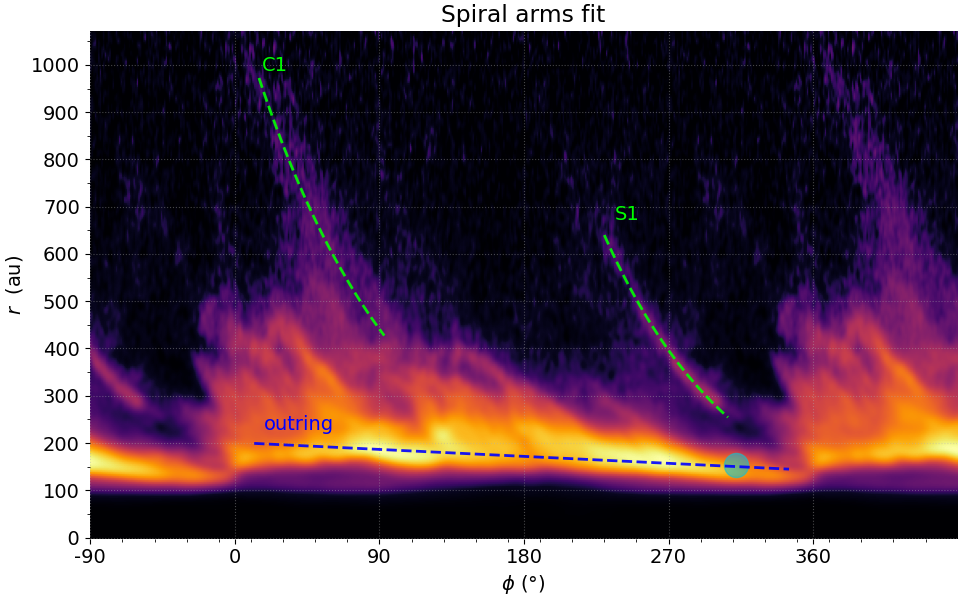}
            \caption{ Polar-coordinate visualisation of the disk of HD\,34700\,A, from IRDIS \Qphi frame, multiplied for a $r^2$ correction factor for the flux dilution and lightly smoothed. Note: the disk is duplicated horizontally to avoid cuts at the image boundaries (notice the $x$ axis). The colour scale is logarithmic to highlight the faint spiral arms. In dashes, the fitted logarithmic (\emph{lime green}) and archimedean (\emph{blue}) spiral arms. The cyan circle around \ang{310} represents the approximate centre of the mm-dust emission. }
            \label{fig:extended_polar}
        \end{figure*}

        The outer spiral arms of HD\,34700\,A have some of the highest pitch angles and radial extents analysed to date \citep[see e.g. Fig. 4e and 4f of][]{Bae+2022}.
        To compare the geometrical properties of the spiral structures, we have produced an extended polar-coordinates visualisation of the disk, shown in \figurename~\ref{fig:extended_polar} and we performed spiral fits to compute the pitch angles. 
        To obtain the $ (r, \phi) $ polar visualisation, we deprojected the IRDIS \Qphi based on the inner ring geometry parameters (see \tablename~\ref{tab:inring}), using the \texttt{scikit-image} module in a custom-made \texttt{Python} code. We did not assume any disk flaring, to avoid imposing a height profile for these outer regions of the disk, where the spiral arms are found, and to be consistent with the scenario outlined in \secname~\ref{subsec:phasefunc}. 

        For the analytical spiral fit we opted to focus on the arms that we were able to distinguish from the rest of the disk. The wealth of substructures in the $ \ang{0} < \phi < \ang{180} $ region, in fact, although interesting, mixes up together in a way that would make any spiral fit partly arbitrary. Qualitatively, however, they appear to have a steep and almost constant slope, as illustrated in \figurename~\ref{fig:extended_polar}. 
        Closely inspecting the deprojected image, we noticed that the bright ridge of the outer ring shows a varying radial distance along the azimuthal dimension, culminating with the discontinuity at \ang{\sim 0}, where there is a sharp jump. In this context, it seemed interesting to calculate the pitch angle of the outer ring, assuming that it was a spiral arm itself as a whole. 
        Then, we decided to fit analytically only the following large-scale features: the outer ring (labelled "outring" for short) and the two long arms extending furthest away from the disk. 
        We retained the name "S1"\footnote{ To be precise, they referred to this arm as "S1b" in their paper, as the continuation of a "S1a" after a shadow. With our polar deprojection these two structures seemed independent of each other; thus, we decided to label S1 the evident log-spiral at $ r > \SI{300}{au} $ and neglect the other fuzzy structure. } 
        for the arm already labelled by \cite{Uyama+2020:HD34700}, and we called "C1" the other arm (as for \emph{Columba}), not to overlap with the previous authors' labels. 
        We sampled the two outer arms picking by hand the coordinates of the brightest parts from \figurename~\ref{fig:extended_polar}. 
        For the outer ring we selected its bright ridge using intensity contours and limiting the azimuthal angles range between $ \ang{10} < \phi < \ang{350} $, to leave a small margin around the discontinuity at \ang{ \sim 0}. 
        Then, we used the \texttt{scipy.curve\_fit} algorithm to fit the collected data points and find the best spiral parameters.         
        We performed the regression with two commonly used types of spiral:
        \begin{equation}
            r(\phi) = 
            \begin{cases}
            a \cdot e^{b \phi}    &   \mathrm{logarithmic} \\
            a + b \phi     &   \mathrm{archimedean} \\
            \end{cases}
        \end{equation}
        where $ (\phi, r) $ are the polar coordinates on the disk plane. 
        We fit the logarithmic shape for the two outer arms and the archimedean for the outer ring, based on their qualitative appearance. Indeed, a logarithmic growth implies a steepening curve in polar coordinates, similarly to the S1 arm; that archimedean spiral, instead, has a constant slope in polar coordinates, more similar to the ridge of the outer ring. Real spiral wakes can take much more complex shapes, however, it is still informative to use these simple shapes to obtain quantitative insights on the disk morphology.  
        
        We report the pitch angles obtained for our spirals in \tablename~\ref{tab:spiralfit}, along with the standard deviation errors. 
        Since the pitch angle of an archimedean spiral is not constant (as is the case for the logarithmic spiral), we chose to report in the table the pitch angle value computed at the smallest radial distance reached by the outer ring (\SI{\sim 120}{\astronomicalunit}). 
        To inspect in greater detail the outer ring pitch angle behaviour, we repeated the analysis with a piecewise fit of archimedean arms. 
        We computed the arm fit parameter on a window of $\Delta \phi =\ang{50}$, running along the outer ring at steps of \ang{5}, to see how its pitch angle evolved with the azimuthal angle. 
        This piecewise computation returns for the outer ring pitch angles within the range between \ang{\sim 1} and \ang{13}, with the highest values near $ \phi \sim \ang{270} $. 
        This could be informative of possible perturbations along a spiral arm. Under appropriate circumstances, and with sufficient resolution, a wake-launching planet can be located where the pitch angle has a cusp, peaking theoretically near \ang{90} values \citep{Zhu+2015:spirals}. No similar feature could be retrieved from our data analysis.

        \begin{table}
            \caption{ Pitch angles of the spiral arms plotted in \figurename~\ref{fig:extended_polar}. The uncertainty for the pitch angles corresponds to 1$\sigma$. The archimedean pitch angle is evaluated at \SI{\sim 120}{\astronomicalunit}. }      
            \label{tab:spiralfit}      
            \centering
            \setlength{\tabcolsep}{12pt}
            \begin{tabular}{lcc }
                \toprule     
                Spiral ID   &   Type   &   Pitch angle (°)     \\
                \midrule 

                C1   &   log  &    $ 31.1 \pm 8.0 $       \\
                S1   &   log   &    $ 34.5 \pm 1.9 $       \\
                outring   &   arch   &   $ 4.4 \pm 0.1 $    \\                       
                \bottomrule 
            \end{tabular}
        \end{table}

        We remark the similarities between C1 and S1, starting from the large pitch angles $\gtrsim \ang{30} $, compatible within uncertainties. 
        The two arms are visualised and plotted on the disk plane in \figurename~\ref{fig:Qphi_r2}, for context.
        They are found on opposite sides of the disk, close to being \ang{180} apart. They are both structures with the largest radial extensions and similar logarithmic shape. S1 is detected up to \SI{\sim 600}{\astronomicalunit}, while C1 extends to the impressive \SI{\sim 1000}{\astronomicalunit} from the central stars.
        The fit of arm C1 has large uncertainties, due to its intrinsic thickness and faintness; also, the presence of the other substructures closer to the outer ring does not allow to constrain its shape at shorter radii. 
        We recall that the existence of these two arms on a single orbital plane is not guaranteed. However, based on our deprojection and the similarities between these spirals, it seems reasonable to assume a connection between them.
        We discuss in further detail the implications of the spiral shapes and pitch angles in \secname~\ref{discuss:spiral_arms} and \secname~\ref{discuss:cloudlet}.

    \subsection{Millimetre-dust asymmetry}
        \label{subsec:mm_dust}

        HD\,34700\,A was observed in the mm wavelengths with ALMA (\figurename~\ref{fig:all_new_wave} and \figurename~\ref{fig:ALMAcont_A+B}). The continuum thermal emission, tracing larger millimetre-dust, appears to be significantly displaced from the centre of the system, where they have been depleted. Similar results were obtained by \cite{Benac+2020:HD34700} using the SMA \citep{Ho+2004:SMA}.
        They argue this could be the manifestation of a dust trap with an asymmetric, crescent-like morphology, where most of the remaining mm grains accumulate. Such asymmetric traps are often (but not exclusively) associated with vortices in the disk \citep[e.g.][]{Birnstiel+2013:dust, Lyra&Lin2013}. 
        
        We traced the contours of the mm emission in the last panel of \figurename~\ref{fig:all_new_wave} and compared it to the disk structures observed in scattered light. 
        The outer parts of the outer ring in the North-West quadrant, the closest to the mm-asymmetry, appear fainter and sparser, perhaps because of a locally lower scale height of the disk (see panels 3 and 6 of \figurename~\ref{fig:all_new_wave} and \figurename~\ref{fig:extended_polar}). 
        Moreover, the mm-dust emission peak is near the point where the outer ring seems to intersect the inner ring, hinting at a perturbed area. It is not centred on the \ang{\sim 0} discontinuity.
        The inner ring, instead, does not appear to be significantly warped in the proximity of the mm emission, but the ring's centre PA is interestingly close to the mm emission PA (see \secname~\ref{subsec:offsets}).
        The simultaneous presence in the disk of HD\,34700\,A of an asymmetric dust structure, in the ALMA continuum, and of spiral arms, seen in visible-IR scattered light, points to a potential connection between these features \citep[see, e.g.,][]{Cazzoletti+2018, VanDerMarel+2021}. 
        We measured the mm-continuum integrated fluxes for HD\,34700\,A and B to be of 9.1\,mJy and 4.0\,mJy respectively, with a rms uncertainty of 0.13\,mJy. The continuum emission around the B component is centred on the star and elliptical, contrary to component A (see \figurename~\ref{fig:ALMAcont_A+B}).

    \subsection{Gas velocity}
        \label{subsec:CO_vel}

        We determined the velocity of emission peak for the \element[][12]{CO} line from the ALMA band 6 spectral datacube, using the \texttt{bettermoments} package \citep{Teague&Foreman2018} to study the gas kinematics. We detect gas rotation patterns for HD\,34700 A and B, whereas we did not detect the C component in either the gas or mm continuum. No gas streamers connecting the A and B components are detected in our data. 
        
\begin{figure}
            \centering      
            \includegraphics[trim=0.45cm 2cm 0  2.5cm ,clip, width=\hsize]{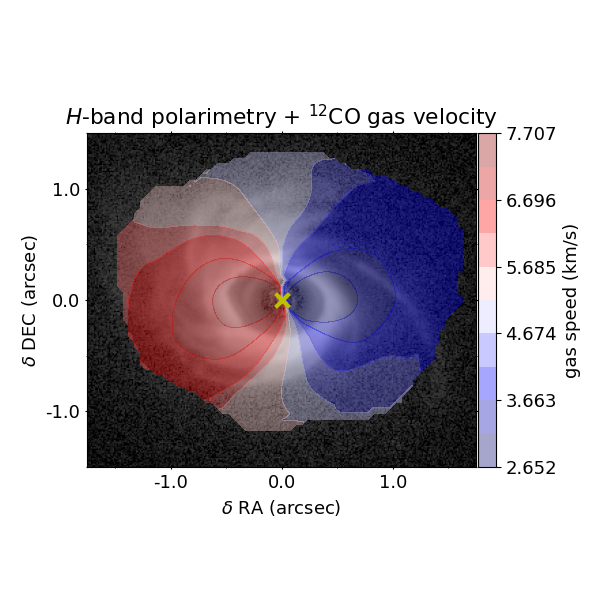}
            \caption{ ALMA data velocity contours for HD\,34700\,A, obtained from the \element[][12]{CO} molecular lines. 
            For reference, we put as background the IRDIS $H-$band \Qphi and marked the source position with a yellow cross.
            The gas shows a well defined rotation pattern, extending to the innermost regions of the disk.  
            The beam is $\ang{;;0.52} \times \ang{;;0.48}$ at a PA of \ang{-61}. The gas speed values refer to the Local Standard of Rest.}
            \label{fig:COmapA}
        \end{figure}

        The velocity map for the A component, shown in \figurename~\ref{fig:COmapA}, displays a rotating gas that extends to the innermost regions of the disk, inside both the outer and inner ring. The rotation is counterclockwise on the sky plane, in agreement with the winding of the spiral arms. 
        The source is well centred in between the two lobes of the rotation pattern, confirming the host star as the barycentre of the system.
        There is no significant velocity perturbation at the location of the mm-dust peak, near the break in the scattered light outer ring, at least at the moderate resolution of these data.
        \figurename~\ref{fig:COmapA} data is masked at $5\sigma$ to avoid including noisy background signal. 
        We estimate for the gas disk a position angle of $ PA_\mathrm{gas} = \ang{93 \pm 2} $, from a straight line fit of the peak velocity coordinates with \texttt{scipy.curve\_fit}.

        The companion HD\,34700\,B was also detected in the ALMA velocity map. The smaller angular size of B, with an equal beam size, implies a worse gas velocity resolution than for A. Nonetheless, we detect a velocity gradient typical of rotating gas in the \element[][12]{CO} line (\figurename~\ref{fig:COmapA+B}).
        The gas and dust emissions are both centred on the source, for B. 
        
        Interestingly, the disk around B rotates clockwise, opposite to the disk around A, and with a very different semi-major axis PA. 
        This misalignment between the two components of HD\,34700 suggests that A and B did not originate from the same coherently-rotating gas disk. Instead, they may have collapsed in different parts of a turbulent fragmented cloud \citep[see e.g.][]{Williams+2014, Luo+2022}. %

    \subsection{Scattering phase functions}
        \label{subsec:phasefunc}

        \subsubsection{Models}
    
        To better characterise the nature of HD\,34700\,A's rings, we extracted their azimuthal brightness profiles and fitted a theoretical scattering phase function. In particular, following a standard approach \citep[e.g.,][]{Engler+2017:polar} we adopted the Henyey-Greenstein phase function \citep{Henyey&Greenstein1941:phasefunc}:
        \begin{equation}
            \label{eq:HGfunction}
            F_{\mathrm{HG}}(\theta, g) \propto \frac{1}{4 \pi} \frac{ 1 - g^2 }{ ( 1 + g^2 - 2 g \cos \theta )^{\nicefrac{3}{2}} } ,
        \end{equation}
        which depends on the scattering angle $\theta$ and the asymmetry parameter $ g $. The latter can take any value in the interval $ [-1,1] $ and parameterises the balance of scattered power between forward $( g>0 ) $ and backward scattering $ ( g<0 ) $. 
        In case of polarised observations (as $ Q_{\phi} $ data) we accounted for the linear polarisation (LP) of light multiplying \eqname~\ref{eq:HGfunction} by the single scattering Rayleigh factor: 
        \[ LP (\theta) \approx  \frac{ 1 - \cos^2 \theta }{ 1 + \cos^2 \theta }  .\]

        A critical uncertainty for the fitting procedure was due to the unconstrained 3D structure of the disk itself, since to provide a scattering angle $ \theta $ for each point of a ring requires assuming a particular geometry, in terms of inclination, PA and flaring. 
        In fact, considering an axially symmetric ring around the central star, observed at an inclination $ i $ and with a locally constant flaring angle $ \gamma $, we can infer the scattering angles following the formulation of \cite{Quanz+2011:scattering} and \cite{Stolker+2016:scattering}:
        \begin{equation}
            \label{eq:thetas}           
            \theta =
            \begin{cases}
                \ang{90} + ( i + \gamma ) \cdot \cos{ \phi} & \quad  \ang{-90} < \phi < \ang{90} \\
                \ang{90} + ( i - \gamma ) \cdot \cos{ \phi} & \quad \ang{90} < \phi < \ang{270}
            \end{cases} , 
        \end{equation}
        
        with $ \phi $ being the azimuthal coordinate on the disk plane, starting from zero at the closest point to the observer. We knew from the literature \citep{Monnier+2019:HD34700, Uyama+2020:HD34700, Benac+2020:HD34700} that the closest side of HD\,34700\,A's disk was in the northern half, so we imposed as zero-point the angle of the semi-minor axis counterclockwise from the North. Then, for a guess of $ \left\{ i, \gamma, \mathrm{PA} \right\} $ as disk geometry, we can map each azimuthal angle of the deprojected ring - and the corresponding median brightness $ I (\phi) $ - to a scattering angle, $ \theta (\phi) $ and thus obtain $ I (\theta) $, in the approximation of single scattering of photons from the star. In other words, $ I (\theta) $ is a measure of the dust brightness seen from the accessible scattering angles. 
        The flaring angle influences also the inferred brightness, when correcting for the dilution of radiation flux (i.e. multiplying for a $ r^2 $ mask from the source position).

        \subsubsection{Procedure}
        
        Based on the previous remarks, the less biased procedure we could use to fit the scattering phase function, was to extract a reasonable $ \left\{ i, \gamma, \mathrm{PA} \right\}_k $ deprojection triplet, via random-sampling, and then run a MCMC regression on the theoretical phase function, to find the best $g$ parameter. The first step fixed the mapping to the median $ I (\theta) $, while the second step found the model parameters minimising the residual to the given data. 
        The best final solution was eventually chosen based on a maximum likelihood estimator, evaluating the overall goodness of fit resulting from the given deprojection geometry and the fitted model parameters. 

\begin{figure}[t]
            \centering
            \includegraphics[trim=0 0 45 30, clip, width=\hsize]{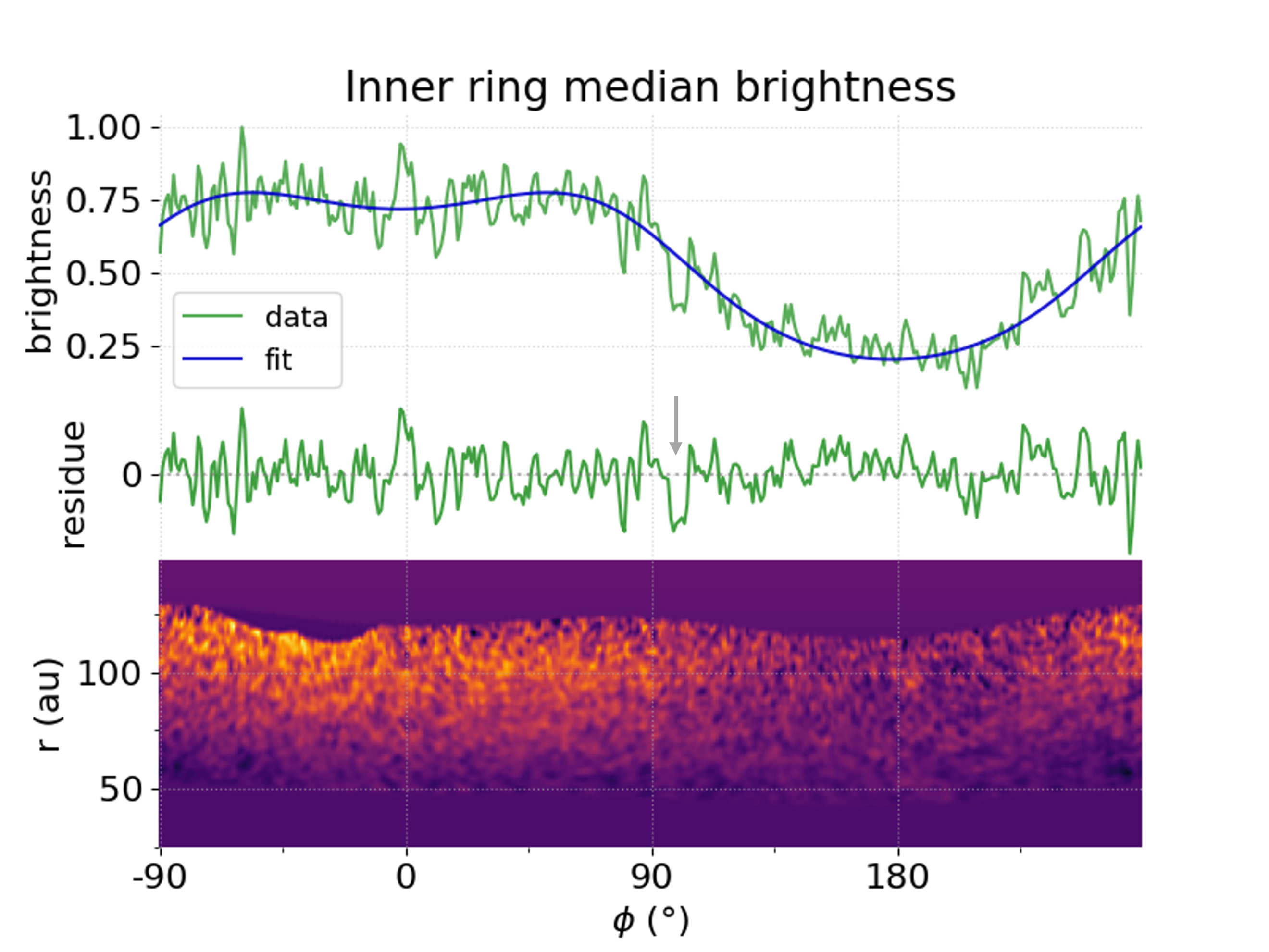}
            \caption{ Analysis of the polarised scattering brightness of the inner ring. \emph{Top:} median radial brightness of the ring (green), as a function of the azimuth angle on the disk plane (\ang{0} is North), normalised to the maximum. 
            \emph{Middle:} residues of the data to the best-fit function, with grey arrows pointing towards the candidate shadow features. \emph{Bottom:} ZIMPOL \Qphi data transformed to polar-coordinates and corrected for the $r^2$ flux dilution, from which the median brightness of the top panel was extracted. The best-fit asymmetry parameter was $ g = 0.41 \pm 0.01 $.}
            \label{fig:inring_profile}
        \end{figure}

\begin{figure}[t]
            \centering
            \includegraphics[trim=0 0 45 30, clip, width=\hsize]{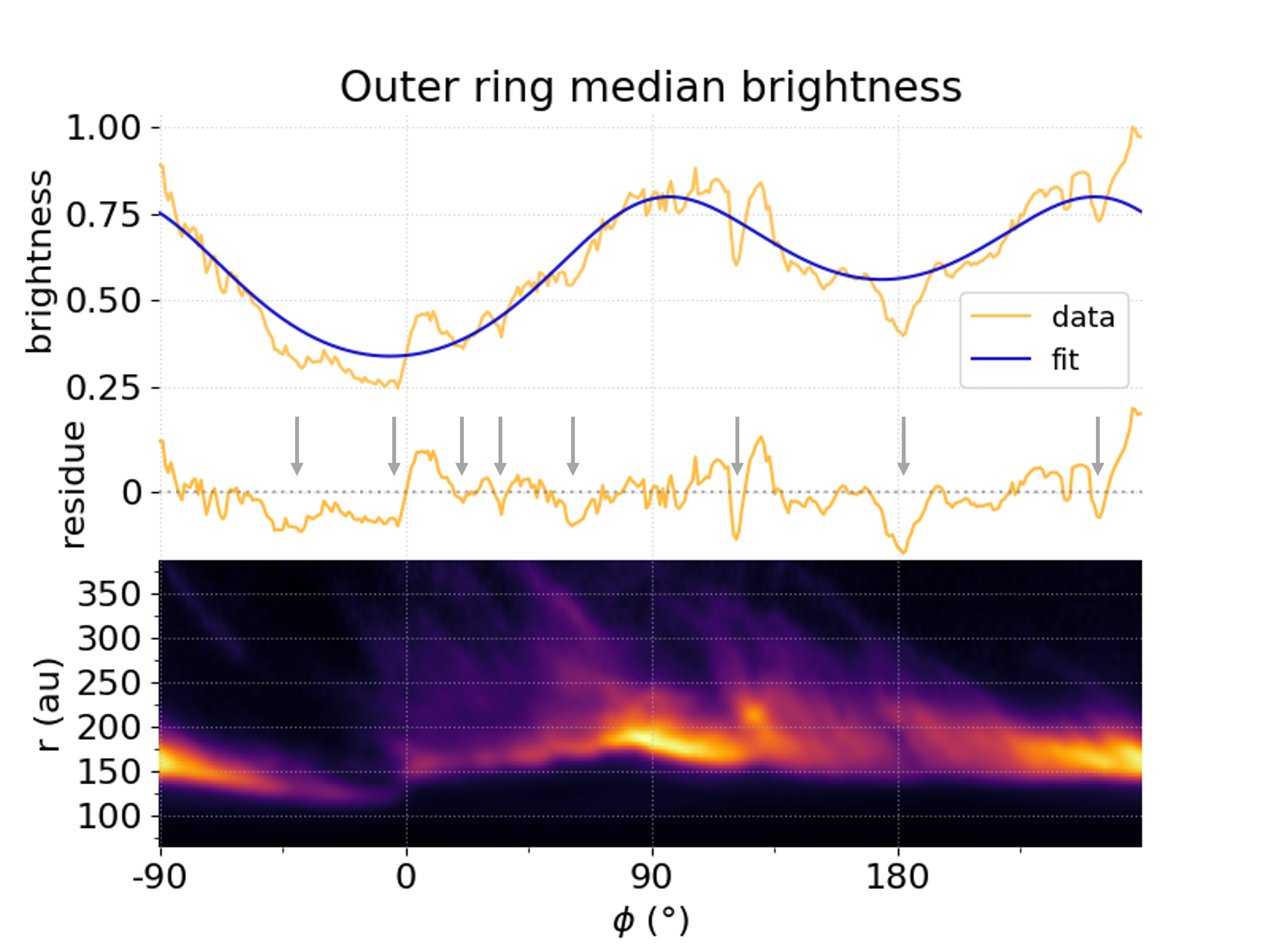}
            \caption{ Analysis of the polarised scattering brightness of the outer ring. \emph{Top:} median radial brightness of the ring (yellow), as a function of the azimuth angle on the disk plane (\ang{0} is North), normalised to the maximum. 
            \emph{Middle:} residues of the data to the best-fit function, with grey arrows pointing towards the candidate shadow features. \emph{Bottom:} IRDIS \Qphi data transformed to polar-coordinates and corrected for the $r^2$ flux dilution, from which the median brightness of the top panel was extracted. The best-fit asymmetry parameter was $ g = -0.16 \pm 0.01 $.}
            \label{fig:outring_profile}
        \end{figure}

        The deprojections were performed with the same method as for the spiral fitting analysis (see \secname~\ref{subsec:spirals}). The polar representations of the rings are shown in the bottom panels of \figurename~\ref{fig:inring_profile} and \figurename~\ref{fig:outring_profile}.        
        The top panels show the normalised brightness of each ring plotted as a function of the disk-plane angles from the North ($\Phi$) in azimuthal slices of \ang{1}. The normalised brightness is measured as the median along the radial dimension. 
        In detail, the inner ring brightness was computed considering intensity contours (on the smoothed data) to find the middle radial distance and taking the median within \SI{30}{au} of thickness around each core point. The outer ring profile, instead, was computed taking the radial median of every pixel within intensity contours chosen to encompass the entire structure, due to its lower regularity; the radial distance was determined based on the maxima for each angle. 
        
        The deprojection parameters were sampled from normal distributions centred on the best fit values from the inner ring geometrical fit (see \tablename~\ref{tab:inring}).
        Once the image had been deprojected, fixing the $ I (\theta)$, a MCMC computation was launched to find the best model parameters for that brightness profile: $ \left\{ g, A, \sigma_\mathrm{j} \right\}_k $, respectively the asymmetry parameter, an amplitude proportionality factor, and a jitter term to accommodate the spread of our data points. 
        We used the \texttt{emcee} module \citep{emcee2013} to perform our MCMC runs, adopting flat priors on the $ g, A, \sigma_\mathrm{j} $ parameters and a standard log-likelihood for a Gaussian distribution of the residues. 
        After a few attempts, we noticed that the flaring angle parameter was giving nonphysical results, warping the rings artificially and pumping the forward scattering part by means of the $ r^2 $ correction. 
        Moreover, estimating the expected flaring based on the offset along the projected semi-minor axis of the disk, we obtained almost negligible angles, in contrast to the MCMC-preferred values (>\ang{15}). This suggested that the scattered light we see does not come from a regular, optically thick surface but could originate from dust at different heights in a thin disk. 
        Motivated by this, we neglected the flaring angle for both rings and launched our MC extractions of $ \left\{ i, \gamma, \mathrm{PA} \right\}_k $ always imposing $ \gamma = 0 $.

        \subsubsection{Results}

        The procedure described above was applied on three independent datasets: the total-intensity iRDI reduction of the IRDIS K-band observation, the IRDIS H-band \Qphi (bottom left panel of \figurename~\ref{fig:IRDIS_quad}) and the ZIMPOL \Ha \Qphi (\figurename~\ref{fig:Qphi_ZIMPOL}) data. The latter, in particular, was masked in two variants, to isolate the inner and outer ring from one another. All of these images allowed an estimate of the outer ring parameters, while only the ZIMPOL \Qphi could be exploited for the inner ring.
        We randomly extracted 100 deprojection triplets for each dataset, to sample the ${i, PA}$ parameter space, and run the MCMC regression on each one. From these runs we selected the best solution as the one yielding the highest likelihood estimator, which assessed the overall quality of the fit. 
        The uncertainties were computed as the $1\sigma$ spread of values on the best fit solution, for $g$. 
        We remark here that the main goal of our phase fitting was to provide an estimate of the asymmetry parameter $g$ and assess the differences between the two rings. 
        A summary of all best-fit results is reported in \tablename~\ref{tab:mcmc_fit}.
     
        The resulting brightness profiles show very different scattering properties between the inner and outer ring. 
        The inner ring, illustrated in \figurename~\ref{fig:inring_profile}, is characterised by a broad forward scattering peak (North is close to the nearest point to the observer) and a weaker back-scattering valley, which translates to a positive asymmetry parameter of $ g_\mathrm{in} = 0.41 \pm 0.01 $. 
        
        The outer ring was analysed in three bands, \Ha, $H$ and $K$. Detailed values for the three bands are reported in \tablename~\ref{tab:mcmc_fit}. 
        The resulting asymmetry parameter values were consistently much smaller for the outer ring than for the inner ring, oscillating around $ g \sim 0 $. This reflects an apparently isotropic scattering of this dust, but it is likely to be an observational effect, as we argue in the Discussion.        
        As shown in the polarimetric observation of \figurename~\ref{fig:outring_profile} there are two brightness peaks, around \ang{\pm 90} away from the closest disk point, and two valleys in between.
        Additional figures illustrating the phase fits on the outer ring for the total intensity data and the \Ha data can be found in the Appendix, in \figurename~\ref{fig:outring_profile_RDI} and \figurename~\ref{fig:outring_profile_ZIMP} respectively. 
        It is particularly striking how, focusing on the North direction (\ang{0} of both rings), the inner ring is close to its maximum brightness, whereas the outer one is in the middle of a valley. This is the cause of a major tension in the value of the asymmetry parameter between the two rings, which we discuss in depth in \secname~\ref{subsec:g_tension}.

        \begin{table}[t]
            \caption{ Best-fit scattering asymmetry parameter $g$ and relative deprojection angles, obtained from the MCMC regression. }             
            \label{tab:mcmc_fit}      
            \centering
            \begin{tabular}{l  C C C}
                \toprule     
                 &   \mathbf  g    &  \mathbf i \ (^\circ)  &  \mathbf{PA} \ (^\circ)  \\ 
                \midrule 
                \midrule
                Inner ring (\Ha, \Qphi )  &   +0.41 \pm 0.01   & 34.09   &    87.72   \\
                \midrule
                Outer ring (\Ha, \Qphi )  &   -0.03 \pm 0.01   & 33.06   &    82.27   \\
                Outer ring  ( H, \Qphi)  &     -0.16 \pm 0.01   &   32.63    &   84.21 \\
                Outer ring (K, $I_{tot}$ )  &   +0.04 \pm 0.01   & 39.29   &    79.00   \\

                \bottomrule    
            \end{tabular}
        \end{table}

    \subsection{Shadowing effects}
        \label{subsec:shadows}
    
        In this section we analyse the polar-coordinates images of the inner and outer disk, looking for effects of shadowing. Similar effects on the outer ring were already detected by \cite{Monnier+2019:HD34700} and \cite{Uyama+2020:HD34700}.

        \textit{Inner ring:} the brightness profile of the inner ring is rather noisy, showing strong fluctuations over angles of few degrees. In this condition, a thin shadow might be buried under this noise or a random fluctuation could be mislabelled as a shadow. When visually inspecting the bottom panel of \figurename~\ref{fig:inring_profile} it is not obvious to identify evidence of shadowing. 
        In the middle panel, where the residuals are reported, we identified only one main dip around \ang{101}, marked by the vertical grey arrow.
        
        \textit{Outer ring:} the outer ring data is less noisy than the inner ring, as fluctuations from one angle to the next are smaller. However, considering its wealth of substructures, it is not trivial to assess the nature of the brightness dips, as real shadows or shapes on the illuminated dust surface. 
        Referring to \figurename~\ref{fig:outring_profile} and starting from the most visually-prominent features, we identified a clear dip near \ang{+182}, one at \ang{+121}, and one near the discontinuity, starting around \ang{-3}. Smaller dips can be found at \ang{+64}, \ang{+20}, \ang{+35} and \ang{254}.
        Then, inspecting also the total intensity data of \figurename~\ref{fig:outring_profile_RDI}, we report another dip at \ang{-34}.
        The brightness profile of the outer ring has so many features that the best fit line itself is heavily influenced by the dips and thus it does not offer an ideal unbiased reference to assess the supposed shadows analytically. 
        
        The alignment of two important dips in the outer ring along the North-South axis, the one at \ang{-3} near the discontinuity and the shadow at \ang{+182}, constitutes an element of interest. 
        Shadowing effects are not rare in protoplanetary disks, and there are several examples of symmetrical shadows at opposite sides of a disk, cast by inner misaligned rings (\citealp[e.g.][]{Benisty+2022} for a review).
        In this case, the shadows have converging angles, as the ones of HD\,100453 \citep{Benisty+2017:hd100453}, GG\,Tau\,A \citep{Keppler+2020:ggTau}, or other disks in Fig.1 of \citet{Benisty+2022}. 
        Instead, in HD\,34700\,A we observe a discontinuity which, if anything, is rather parallel to the shadow at \ang{183} and, maybe not coincidentally, is also parallel to most spiral substructures. 
        Neglecting for a moment the shadow angles, we can wonder which disk structure is projecting its shadow onto the outer ring. 
        The inner ring resolved by ZIMPOL, based on our analysis, appears sufficiently aligned with the outer ring that it would be difficult to create a thin shadow such as the one at \ang{182}. Most importantly, this shadow is near the disk minor axis, whereas from geometrical considerations it should be projected along the major axis (except for extreme disk warps, not supported by our data). 
        Another hypothesis could be an unresolved and misaligned ring close to the central binary, launching its shadow on both the inner and the outer rings. However, we report no detection of a \ang{\sim +182} shadow in the ZIMPOL's inner ring (\figurename~\ref{fig:inring_profile}) and in general no correlation with any other brightness dip found on the outer ring. 
        As a consequence, we suggest that the dip at \ang{+182} may be linked to the shapes of the disk and spiral arms, instead of being a projected shadow. It could be similar to the dip at \ang{+121}, with the main difference that the first one is better resolved, thanks to a lucky geometrical alignment.

    \subsection{Mass constraints on potential detectable companions}
        \label{sec:detect_limits}

        \begin{figure}
            \centering
            \includegraphics[width=\hsize]{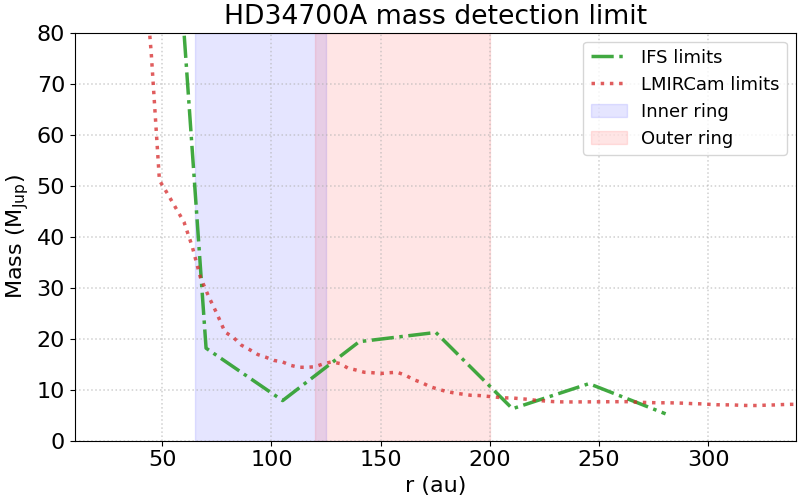}
            \caption{ Limits on the mass of a planetary companion in the disk of HD\,34700\,A, based on the reduced observations from IFS (\emph{green line}), LMIRCam (\emph{red dotted line}) and the AMES-COND models for a young gas giant \SI{5}{Myr} old. Shaded areas represent the rough radial extension of the inner ring (\emph{blue}) and the outer ring (\emph{red}).}
            \label{fig:detect_limits}
        \end{figure}

        The new SPHERE/IFS and LBT/LMIRCam images allowed us to put constraints on the mass limit of a possible candidate companion.   
        We calculate these limits using the method described in \cite{Mesa+2015:IFS} and show them in \figurename~\ref{fig:detect_limits}. The luminosity contrasts are converted into mass detection limits based on the "AMES-COND" theoretical models for the atmospheres of non-irradiated young gaseous planets/brown dwarfs with no dust opacity \citep{Allard+2001}, and assuming a system age of \SI{5}{Myr}. 

        The presence of the bright outer ring worsens the detectability of substellar objects in its proximity (light red shaded area in \figurename~\ref{fig:detect_limits}), whereas the inner ring has a negligible impact at the observed wavelengths. The L' band LMIRCam data improve the IFS detection limit in the outer ring region, pushing it down to 10-\SI{15}{\Mjup}, and for small separations (\SI{<60}{\astronomicalunit}), where only massive BDs are detectable anyway. From IFS, we have a small window around \SI{100}{au} from the star, where we should be able to spot giant candidate companions of \SI{10}{\Mjup} and then outward \SI{200}{au} again. 
        Further than \SI{280}{au} we should be able to detect gas giants of the order of \SI{5}{\Mjup}, but we did not find any. The mass limits we obtain are compatible with those of \cite{Uyama+2020:HD34700} for the $JHK$ bands.

\section{Discussion}
    \label{sec:discuss}

    HD\,34700\,A circumbinary disk mainly consists of two very different ring-like substructures, whose morphology we have analysed in detail in \secname~\ref{sec:analysis}. 
    The inner ring centre is offset with respect to the star, which suggests that it could have a non-null eccentricity. 
    The outer ring-like dust structure is tied to wide spiral arms with similar logarithmic shape. 
    The two ring structures do not show equal scattering brightness in all passbands: the inner dust appears qualitatively bluer than the outer one, as it is not detected in the longer-wavelength near-infrared data. 
    In addition, their scattering phase functions are fit to significantly different effective asymmetry parameters. 
    Almost all mm-continuum emission in HD\,34700\,A is asymmetric and centred away from the host star. 
    We detect CO gas rotation in components A and B, but no gas streamer in between, as reported in \secname~\ref{subsec:CO_vel}. 
    
    In the following, we discuss plausible physical causes of this diversity in the disk of HD\,34700\,A. Based on our results and on previous literature, we attempt to match the various pieces together, to create a globally coherent model of this system and assess its limits.

    \subsection{ Scattering asymmetry parameter tension}
        \label{subsec:g_tension}

        The scattering phase functions of the inner and outer ring around HD\,34700\,A reveal evidence of the diversity in the dust populations of the two rings. 
        Their asymmetry parameters $g$, defined and calculated in \secname~\ref{subsec:phasefunc}, are radically different. 
        On the one hand, the $ g \approx 0 $ value fit for the outer ring would suggest that its dust is scattering isotropically. This generally happens in the Rayleigh regime, when the grains are much smaller than the wavelength. 
        On the other hand, the inner ring fit returned $ g \sim 0.4 $, which would reflect more forward scattering, typical of grains with size comparable to the scattered wavelength.         
        However, the scattering cross section of dust aggregates drops rapidly for wavelengths longer than their size \citep[see e.g.][]{Tazaki&Tanaka2018:dustscat, Min+2016:dust}. 
        This implies that, for a given filter wavelength $ \lambda $, we expect to see particles of size $ a \gtrsim \lambda $, where $ a $ is the characteristic radius. The outer ring is perfectly detected up to the $K2$ band. Assuming that the detection limits for ZIMPOL in \Ha are not better than for IFS and IRDIS bands, the inner ring is properly detected only in \Ha, suggesting a qualitatively smaller grain size. 
        Then, our results appear in contradiction: the outer ring should not display isotropic scattering and at the same time consist of particles larger than those in the inner ring. 
        This tension can be solved taking into account the difference between the theoretical scattering pattern and the one actually measured from observations.
        
        \cite{Min+2016:dust} derived the optical properties of dust aggregates with detailed numerical computations using the discrete dipole approximation, providing values of the asymmetry parameter as a function of wavelength and for various grain sizes.
        They find that the \emph{effective} asymmetry parameter, which is obtained by fitting observations of a limited range of scattering angles, can have very different values than the \emph{formal} asymmetry parameter, resulting from the complete theoretical computation of the aggregate properties. 
        Comparing their Fig. 5 \citep{Min+2016:dust} with our results, we find that for the \Ha band ($\lambda \sim 0.65\,\mu$m) and for particles with size $a = \SI{0.4}{\um}$ ($r_\mathrm{V}$ in their paper), their effective $g$ parameter is very close to our fit value. To find an effective $g \sim 0$ value in their results, the particle size increases to $ a = \SI{4}{\um} $. And for these particles, the effective $g$ is close to zero for wavelengths going from \Ha to $K2$, which is perfectly in line with our results for the outer ring, across the different datasets we analysed. 
        Interestingly, for \SI{4}{\um} grains the corresponding formal asymmetry parameter is around 0.8: the scattering is not at all isotropic, it is strongly forward scattering. The effective parameter is much lower exactly because a great amount of light is scattered to small angles that are not accessible by our observation (same as \citealp{Min+2016:dust} for a disk inclined of \ang{\sim 40}). The difference between effective and formal asymmetry parameter should be taken into account when fitting from observations. 
        \cite{Min+2016:dust} conclude that the $ g $ parameter is almost only influenced by the size of the aggregate as a whole, making it a good size indicator. 
        This allows us to reconcile the results of the phase function fitting with the qualitative expectations based on the ring detection wavelengths. Thus, we assume that the outer ring is mainly composed of grains with $ a_\mathrm{out} \gtrsim \SI{4}{\um} $, while the inner ring consists of grains with $ a_\mathrm{in} \lesssim \SI{0.4}{\um} $.

        A possible caveat to the inner ring $g$ value is that an azimuthally uniform dust distribution was assumed. A similar brightness variation could in theory be reproduced by a non-axisymmetric dust ring: the brighter North side could have a higher particles density, which can be misinterpreted with a stronger forward scattering component. 
        Finally, we remark again that the ultimate goal of the phase function fitting was not to constrain the parameter $g$ itself, but to find a proxy for the grain size differences between inner and outer ring. Constraining the real value of the asymmetry parameter $g$ is a tricky task, based on single-angle observations, and depends on the adopted scattering phase function, together with the assumptions on grain size distribution.

    \subsection{A challenging dust segregation}
        \label{subsec:dust_segreg}
        
        The grain size difference, inferred from the scattering phase functions and discussed in \secname~\ref{subsec:g_tension}, requires a physical mechanism responsible for the observed segregation.
        Dust segregation might be caused by a giant planet creating a pressure bump outside its orbit and the gap between the two rings. A pressure bump acts as a filter for larger grains, preventing their inward migration, while allowing smaller dust particles to filter through and populate the inner regions \citep{Rice+2006:dust, Zhu+2012:dust, Pinilla+2012b:dust, Pinilla+2016:dust}. 
        However, this filtering is more efficient for particles close to millimetre size and above, since smaller dust has a stronger diffusion and is more coupled to the gas component. The depth of the gap increases with the mass of the planet, filtering out progressively smaller dust grains \citep{Zhu+2012:dust}. 
        The dust size deduced from our observations of HD\,34700\,A is approximately two orders of magnitude smaller than the thresholds of \cite{Zhu+2012:dust} for effective filtering, as a general reference. Due to turbulent diffusion, in fact, they found that even a \SI{6}{M_{J}} planet can only segregate dust bigger than \SI{100}{\um}. 
        
        The presence of more massive planets between the system's rings would be discouraged by the contrast estimates of both this work (\secname~\ref{sec:detect_limits}) and \cite{Uyama+2020:HD34700}, but cannot be excluded.
        The limiting contrast, as a matter of fact, provides the minimum planetary mass that can be seen in front of the local background (actually averaged on circular annuli).
        If, for any reason, a compact object is embedded into a dust structure, the contrast limits may be misleading, as the emitted radiation is extincted within the dust cloud, with sufficient optical depth.
        In this regard, then we must admit the difficulty of ruling out the presence of planets with masses $ \gtrsim 6\,\Mjup$ in a complex system as HD\,34700\,A.
        For example, the mm-dust structure seen from the mm-data might be hiding a compact object, whose light could be completely absorbed by the dust around it. This object would be more massive than our limiting contrasts foresee. 
        Having said that, the possibility that the mm-dust substructure consists of a vortex with no compact object inside is also valid \citep[e.g.,][]{Zhu+2014:dusttrap}.
        Longer-wavelength observations, using e.g. the JWST Mid-Infrared Instrument (MIRI), might help reduce the extinction from the dust around a hypothetical compact object and better constraining its mass.      
        Anyway, in terms of dust segregation, consistently with the studies of \cite{Zhu+2012:dust}, \cite{Pinilla+2012b:dust}, the observed mm-dust structure retains much larger grains than those revealed by IRDIS images in the outer ring. The latter, in fact, does not seem dramatically altered by the trap. 
        
        Lastly, if dust filtration is not effective for micrometre-sized grains, the difference in composition between the two rings could be regulated by growth/fragmentation and radial drift dynamics at the two sides of the gap. 
        In particular, starting from an ideally homogeneous grain size distribution in both rings, we can argue that the presence of the gap, although it does not prevent the filtration of \si{\um}-dust from outer to inner ring, enhances the efficiency of grain growth near the pressure bump \citep{Pinilla+2012b:dust}. This would result in an increased size of the dust observed in the outer ring (\SI{\geq 4}{\um}) compared to the inner (\SI{\leq 0.4}{\um}). In the inner ring, higher particle speed could lead to increased fragmentation rates.
        In addition, the inner ring may also be depleted of "larger" grains ($\gtrsim \SI{1}{\um}$), because of the radial drift. 
        The drift speed, for a given grain size, depends on the local gas density, which sets the Stokes number \citep[see e.g.][]{Birnstiel&Andrews2014, Andrews2015:ppds}. 
        For HD\,34700\,A, then, the grains $\gtrsim \SI{1}{\um}$ may be depleted from the inner ring much faster than those $ < \SI{1}{\um}$ and also faster than their growth timescales. Eventually, the inner ring would be left only with the smallest grains, $\leq \SI{0.4}{\um}$. 
        Detailed theoretical simulations of this disk are required to settle this argument quantitatively, taking into account also the effects of vertical dust distribution. We save a similar approach for future studies.

    \subsection{Spiral arms excitation}
        \label{discuss:spiral_arms}

        In \secname~\ref{subsec:spirals} we analysed the spiral structures observed in the disk of HD\,34700\,A. 
        The literature on spiral arms is solid on the theoretical side, where simulations are able to produce spiral perturbations as a result of several different physical causes, such as planets \citep{Zhu+2015:spirals, Zhu&Zhang2022:spirals}, flybys \citep{Smallwood+2023, Cuello+2023}, binarity \citep{Calcino+2023}, gravitational instability \citep{Forgan+2018b} and vortices \citep{Paardekooper+2010, Huang+2019:vortex}. 
        Observationally, however, it is difficult to determine the cause of a given spiral wake, as the spiral morphology can be degenerate \citep[see, e.g.,][]{Forgan+2018b, Bae+2022}. 
        In general, the presence of a compact object in a disk is manifested through inner and outer spiral wakes. The outer wakes have a vanishing pitch angle with increasing radius. With enough spatial resolution, the pitch angle is expected to peak near the position of the perturber, theoretically reaching values near \ang{90}, with heavier objects yielding larger angles \citep{Zhu+2015:spirals}. 
        Real spiral wakes most often do not adhere to a simple analytical shape. The spiral fitting is relevant in the context of describing a qualitative overall morphology and providing the pitch angle's behaviour. 
        Addressing first the outer ring itself: the pitch angle is higher at smaller radii, in the $ \ang{270} < \phi < \ang{360} $ quadrant, but is limited to moderate values (\ang{\leq 13}). 
        The substructures above the outer ring (not fitted), assuming they are spiral arms, show a much higher pitch angle. \citet{Zhu&Zhang2022:spirals} demonstrated how eccentric massive planets can have a strong impact on the shape of spiral wakes, creating complex overlapping patterns. They themselves mention that HD\,34700\,A's spirals would be naturally produced by an eccentric perturber.

        Concerning arms C1 and S1, they stand out from the rest of the disk for their extension and their large pitch angles $ \gtrsim \ang{30}$. 
        Based on the similarities we found between C1 and S1, it seems reasonable to assume a physical connection between them. 
        Overall, their growing slope with increasing radius would point to a perturber outside the observed disk, such as a bound compact object or a flyby event.
        \cite{Uyama+2020:HD34700} already suggested a flyby between HD\,34700 A and B components, which should have been closer in the past 8000 years, as they infer from proper motion calculations. 
        On the one hand, simulations of \citet{Cuello+2020:ppds} and \citet{Cuello+2023} show that flybys indeed produce large scale regular spiral patterns with large pitch angles. On the other hand, they found that these induced spiral arms lasted only for a few dynamical times after the flyby (which happens only once), translating to a typical dissipation time of a few thousand years. \SI{8000}{yr} could then push the upper limit on survival time, although not ruling it out. 
        On a side note, if the flyby is close enough, the perturber can steal some dust and create a disk of its own \citep{Cuello+2023}. If HD\,34700\,B was the flyby perturber and has the same age as the C component (as expected), this could explain why we do see a mm-dust disk around B and not around C, in the ALMA continuum data.
        One last hypothesis for the creation of C1 and S1 spirals involves the possibility of accreting new interstellar matter from cold gas clouds, which we illustrate separately in the next section.

    \subsection{Cloudlet capture and late infall}
        \label{discuss:cloudlet}

        At a crossroad between dust segregation and ring morphology, we mention the possibility of a late accretion event \citep[see e.g.][]{Ginski+2021:SUaur, Mesa+2022:DRtau}. 
        It is possible for a young star moving through a heterogeneous formation environment (giant molecular clouds) to encounter condensations of cold insterstellar gas, or \emph{cloudlets} \citep{Dullemond+2019}. 
        Whether the primordial disk is already depleted or is still present, the freshly captured gas will evolve to form a new disk around the central star.
        Referring to the work of \cite{Kuffmeier+2020}, we can find similarities between the appearance of the newly formed disk and HD\,34700\,A. In particular, the simulations illustrated in Fig.6 of \cite{Kuffmeier+2020}, display a wealth of spiral arms due to the infall of the cloudlet gas, creating a "flocculent" new disk that qualitatively resembles the outer ring of HD\,34700\,A.
        Moreover, peculiar to cloud capture is the creation of a widely extended and sparse spiral arm, connecting the original gas cloud to the material condensing around the star (as shown in the different simulations of \citealt{Kuffmeier+2020}). 
        The spiral arm C1 analysed in \secname~\ref{subsec:spirals}, may result from a similar cloudlet capture event.
        We might also interpret the HD\,34700\,A inner ring as the remains of a primordial disk (ALMA data point to a high depletion of mm-sized grains in the disk) and the outer ring and spirals as the results of a late infall event. This hypothesis might provide a reasonable justification for the wildly different morphology of the two rings and the properties of their dust grains.
        Some caveats of this argument concern the intrinsic properties of the interstellar dust captured, which should be more pristine than the processed material of the original disk, and the interaction between a pre-existing disk and the captured gas. Considering statistically random orientations for the star-cloudlet encounter and for their angular momenta, the creation of a new ring lying on the same plane as the pre-existing disk, without disrupting it, might be an unlikely coincidence. 
        Lastly, the outermost shells of the gas cloud can sometimes be detected as extended reflection nebulosity near the star, as for the case of FU Orionis \citep{Dullemond+2019} or SU Aur \citep{Ginski+2021:SUaur}.
        We did not observe similar features around HD\,34700\,A, but the capture conditions are relevant in this regard and may not apply to our case. 
        Overall, the late infall scenario cannot be ruled out.

\section{Conclusions}
    \label{sec:conclude}

    We have reported the discovery of a new ring within the disk of the binary star HD\,34700\,A. This new inner ring was detected thanks to SPHERE/ZIMPOL instrument in the \Ha band, at shorter wavelengths than the already known outer ring (which is bright also in the $J$, $H$ and $K$ bands). 
    The variety of substructures in this disk suggests the presence of several disk-shaping mechanisms acting simultaneously. We argue that the disk inner features (ring shapes, gaps, dust segregation, mm-dust asymmetry) may be shaped by a not-yet-seen companion of a few Jupiter masses, whereas the outer disk morphology and the extended spirals may be created by either cloudlet capture or flybys. 
    We summarise in the following the key findings of this work:
    \begin{itemize}
        \item We resolved an inner ring extending approximately from \SI{65}{\astronomicalunit} to \SI{120}{\astronomicalunit}, with a $PA \approx \ang{87} $. It is likely to be a slightly eccentric ring, based on the centre offset with respect to the host stars. 
        
        \item The LoS inclination inferred for the inner ring (\ang{\sim40}) is compatible with those of the outer ring and of the binary host (with the caveat of unknown PA for this). This suggests a strong coplanarity of the whole system, up to large radial separations of hundreds of au. 

        \item We find a sharp segregation in the dust populations of the two rings. 
        Based on the rings detection passbands and the effective scattering asymmetry parameters, the typical grain sizes may differ by one order of magnitude between the inner and outer ring, which may then consist of aggregates roughly of $ \lesssim \SI{0.4}{\um} $ and $ \gtrsim \SI{4}{\um} $, respectively. 

        \item We detected two very extended spiral arms with similar shapes and high pitch angle ($\gtrsim \ang{ 30}$), at opposite sides of the disk. We cannot rule out either the flyby or the cloud capture scenarios as potential excitation mechanisms.
        
        \item The outer ring appears as a spiral arm itself, with a varying radial distance from the star and complex substructures. It is not clear whether the two rings are spatially contiguous. 

        \item We obtained the CO velocity map for HD\,34700\,B and find that the latter rotates with a completely different orientation than the A component. No CO or mm-dust streamer could be resolved between A and B, from our ALMA data.
    \end{itemize}
       
    The mechanisms responsible for dust segregation still require further investigation, but there are hints of intense hydrodynamical perturbations in the disk of HD\,34700\,A. 
    One or more compact objects with several Jupiter masses could be responsible for carving the gap between the rings, creating the mm-dust crescent and possibly segregating the dust grain sizes to some extent.    
    The dust inward migration, diffusion, and accumulation in the pressure maxima could have a particular equilibrium, giving rise to the differentiated rings of this system. 
    In addition, the inner ring cavity is too large to be carved by the tight binary star, so other processes must be at work to explain that.   
    We cannot rule out the presence of substellar companions, based on our detection limits, but future observations with greater sensitivity instruments (JWST or the future ELT) may hold the key.
    Alternatively, higher spectral and spatial resolution ALMA data may find kinematic anomalies within the disk, or \emph{velocity kinks}, as described by \cite{Perez+2015}. This technique has already detected several potential signatures of giant planets and can be used also for embedded bodies \citep[see, e.g.,][]{Teague+2018, Pinte+2020, Pinte+2022}.
    We save for future studies the further characterisation of some features in the disk of HD\,34700\,A which were not yet conclusively framed and that could be key elements for creating a complete physical model of this system.


\begin{acknowledgements}
    This work has been partially supported by the Large Grant INAF 2022 YODA (YSOs Outflows, Disks and Accretion: towards a global framework for the evolution of planet forming systems). 
    The LBT is an international collaboration among institutions in the United States, Italy and Germany. LBT Corporation partners are: The University of Arizona on behalf of the Arizona Board of Regents; Istituto Nazionale di Astrofisica, Italy; LBT Beteiligungsgesellschaft, Germany, representing the Max-Planck Society, The Leibniz Institute for Astrophysics Potsdam, and Heidelberg University; The Ohio State University, representing OSU, University of Notre Dame, University of Minnesota and University of Virginia.
    We acknowledge the use of the Large Binocular Telescope Interferometer (LBTI) and the support from the LBTI team, specifically from Amali Vaz, Jared Carlson, Jennifer Power, Steve Ertel.
    This publication makes use of VOSA, developed under the Spanish Virtual Observatory project supported from the Spanish MICINN through grant AyA2008-02156.
    This work has made use of data from the European Space Agency (ESA) mission {\it Gaia} (\url{https://www.cosmos.esa.int/gaia}), processed by the {\it Gaia} Data Processing and Analysis Consortium (DPAC, \url{https://www.cosmos.esa.int/web/gaia/dpac/consortium}). Funding for the DPAC has been provided by national institutions, in particular the institutions participating in the {\it Gaia} Multilateral Agreement.
    T.B. acknowledges funding from the European Research Council (ERC) under the European Union's Horizon 2020 research and innovation programme under grant agreement No 714769 and funding by the Deutsche Forschungsgemeinschaft (DFG, German Research Foundation) under grants 361140270, 325594231, and Germany's Excellence Strategy - EXC-2094 - 390783311.
    S.F. is funded by the European Union under the European Union’s Horizon Europe Research \& Innovation Programme 101076613 (UNVEIL).
    Ch. Rab is grateful for support from the Max Planck Society and acknowledges funding by the Deutsche Forschungsgemeinschaft (DFG, German Research Foundation) - 325594231.
    A.R. has been supported by the UK Science and Technology research Council (STFC) via the consolidated grant ST/W000997/1 and by the European Union’s Horizon 2020 research and innovation programme under the Marie Sklodowska-Curie grant agreement No. 823823 (RISE DUSTBUSTERS project). 
    This project has received funding from the European Research Council (ERC) under the European Union's Horizon 2020 research and innovation programme (PROTOPLANETS, grant agreement No.~101002188), as well as under the Horizon Europe research and innovation program (Dust2Planets, grant agreement No. 101053020). 
    A.Z. acknowledges support from ANID -- Millennium Science Initiative Program -- Center Code NCN2021\_080
\end{acknowledgements}

\bibliographystyle{aa}
\bibliography{ExoDisks_Biblio}

\begin{appendix}

    \section{Additional figures} 
        \label{appendix:SPF}

        We add to this appendix a few extra figures to complement the already numerous figure within the text. 

\begin{figure*}[hb!]
            \centering
            \includegraphics[width=0.8\hsize]{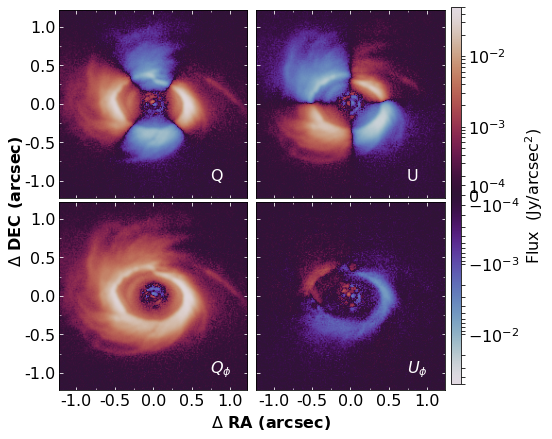}
            \caption{ Stokes parameters frames for the polarimetric observation of HD\,34700\,A with IRDIS in the $H$ band. The subplots are, in order: the $Q$, $U$, \Qphi and $U_\phi$ frames, as overlaid in white text. The colour scale, shared between all four frames, is a symmetric logarithmic function, chosen to display all the fainter details. The coordinates are reported as delta offsets with respect to the nominal position of the central star. }
            \label{fig:IRDIS_quad}
        \end{figure*}

        In \figurename~\ref{fig:IRDIS_quad} we show for completeness the Stokes parameters frames for the IRDIS H-band data of 2019. These have been illustrated with a composite logarithmic colour scale that helps to highlight the faint features of the frames.

        In \figurename~\ref{fig:outring_profile_RDI} and \figurename~\ref{fig:outring_profile_ZIMP} we show the results of the phase fitting procedure on other two datasets for the outer ring(in addition to the one shown within the main text). These allow one to form a more complete panoramic of the scattering properties of the outer dust, and extend the wavelength domain where this study was performed. 

\begin{figure}      
            \centering
            \includegraphics[width=\hsize]{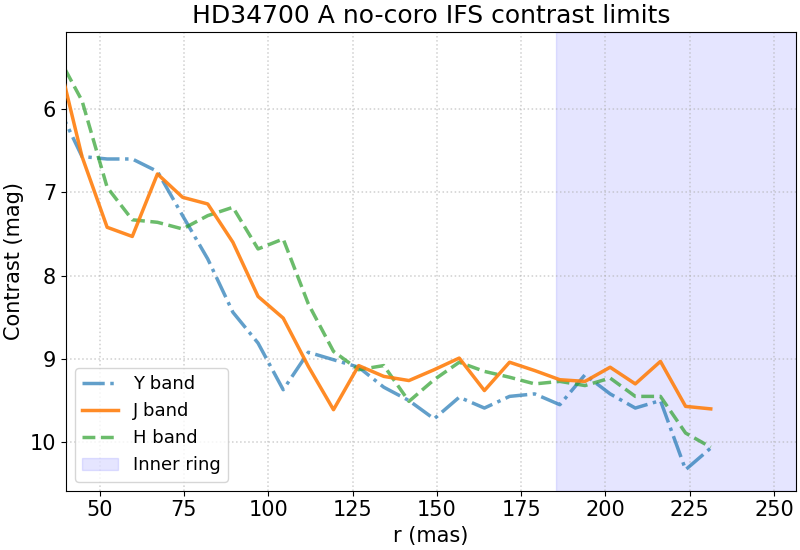}
            \caption{ Contrast limit curves as a function of the distance from the central star HD\,34700\,A, for the inner system region (see \figurename~\ref{fig:nocoro_map}). The blue shaded area represents the inner ring (partially), while the coloured lines the contrasts for the three IFS broad bands (see plot legend). These were obtained from the IFS flux calibration frames, with no coronagraph. Below \SI{42}{mas} self-subtraction becomes more significant and thus this range was excluded. }
            \label{fig:nocoro_lims}
        \end{figure}

\begin{figure}      
            \centering
            \includegraphics[width=\hsize]{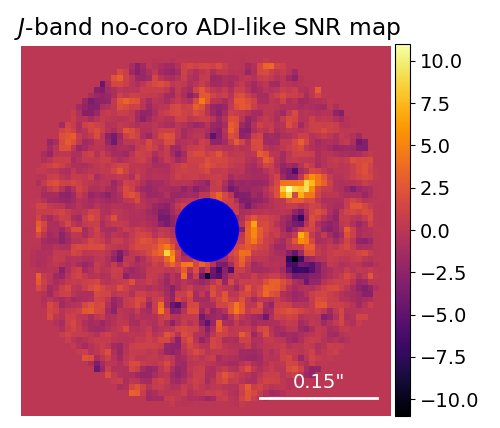}
            \caption{ Map of the signal-to-noise ratio, relative to the ADI-like reduction described in \secname~\ref{subsec:psfsearch} on the 26-11-2021 IFS observation of HD\,34700\,A. The central blue circle masks the area within 1.3 $\lambda / D $, where self-subtraction is significant. The colourbar indicates the SNR level. A couple of higher-SNR points on the Western side may be spurious speckles; no reliable point-source is detected in this reduction. North is up and East to the left. }
            \label{fig:nocoro_map}
        \end{figure}

\begin{figure}      
            \centering
            \includegraphics[trim=0 0 45 30, clip, width=\hsize]{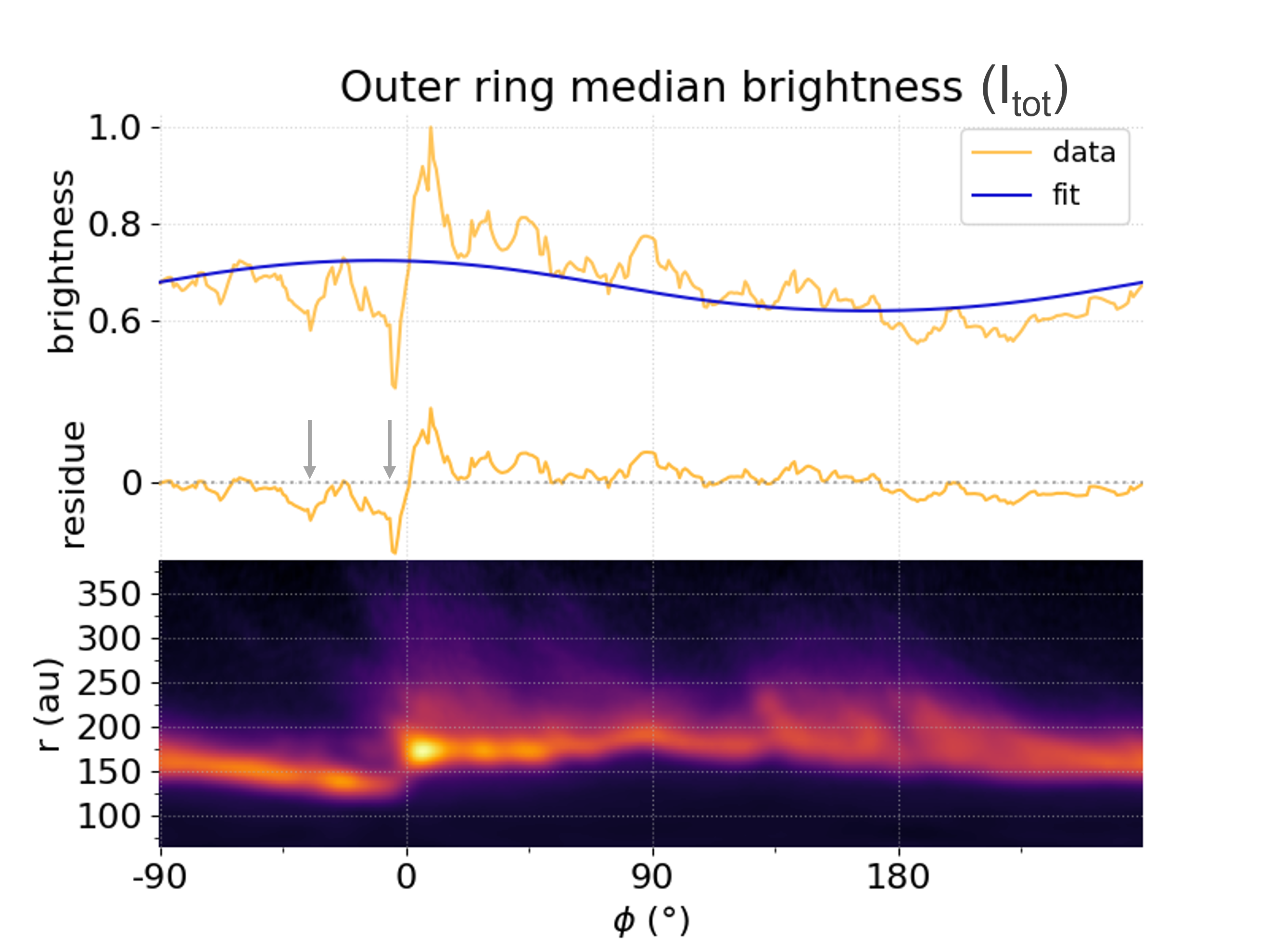}
            \caption{ Analysis of the total intensity scattering brightness of the outer ring on the iRDI K band data. \emph{Top:} median radial brightness of the ring (yellow), as a function of the azimuth angle on the disk plane (\ang{0} is North), normalised to the maximum. 
            \emph{Middle:} residues of the data to the best fit function, with grey arrows pointing towards the candidate shadow features. \emph{Bottom:} IRDIS \Qphi data transformed to polar-coordinates and corrected for the $r^2$ flux dilution, from which the median brightness of the top panel was extracted. The best fit asymmetry parameter was $ g = +0.04 \pm 0.01 $.}
            \label{fig:outring_profile_RDI}
        \end{figure}

\begin{figure}
            \centering
            \includegraphics[trim=0 0 45 30, clip, width=\hsize]{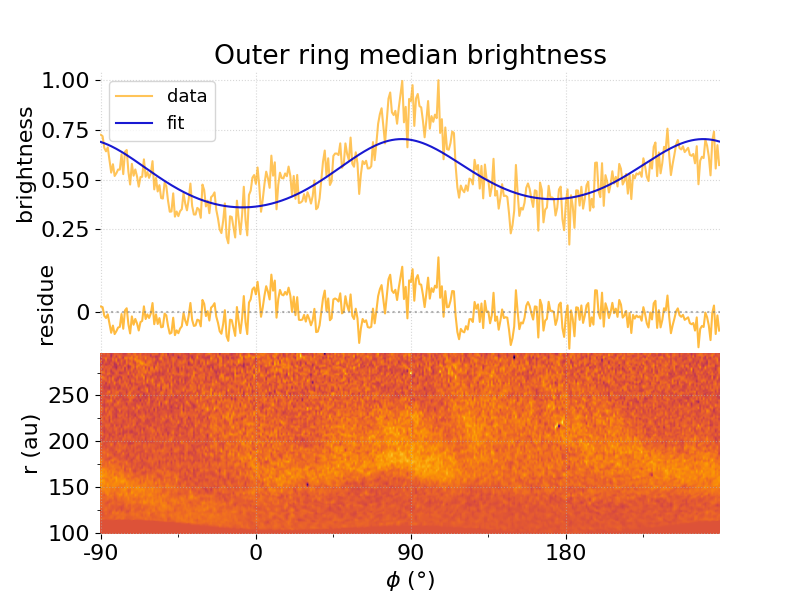}
            \caption{ Analysis of the polarised scattering brightness of the outer ring on the ZIMPOL data. \emph{Top:} median radial brightness of the ring (yellow), as a function of the azimuth angle on the disk plane (\ang{0} is North), normalised to the maximum. 
            \emph{Middle:} residues of the data to the best fit function. \emph{Bottom:} IRDIS \Qphi data transformed to polar-coordinates and corrected for the $r^2$ flux dilution, from which the median brightness of the top panel was extracted. The best fit asymmetry parameter was $ g = -0.03 \pm 0.01 $.}
            \label{fig:outring_profile_ZIMP}
        \end{figure}

        We show the contrast limit curve for the IFS non-coronagraphic data reduction of the 26-11-2021 in \figurename~\ref{fig:nocoro_lims}. The relative signal-to-noise image map is shown in \figurename~\ref{fig:nocoro_map}. These allowed us to search for stellar companions in the vicinity of the central binary, as described in \secname~\ref{subsec:psfsearch}. 
        The limits obtained from the 14-11-2021 data were of worse quality, and we did not show them. 

\begin{figure*}[hbt]
            \centering
            \includegraphics[trim=50 60 0 60, clip,width=\hsize]{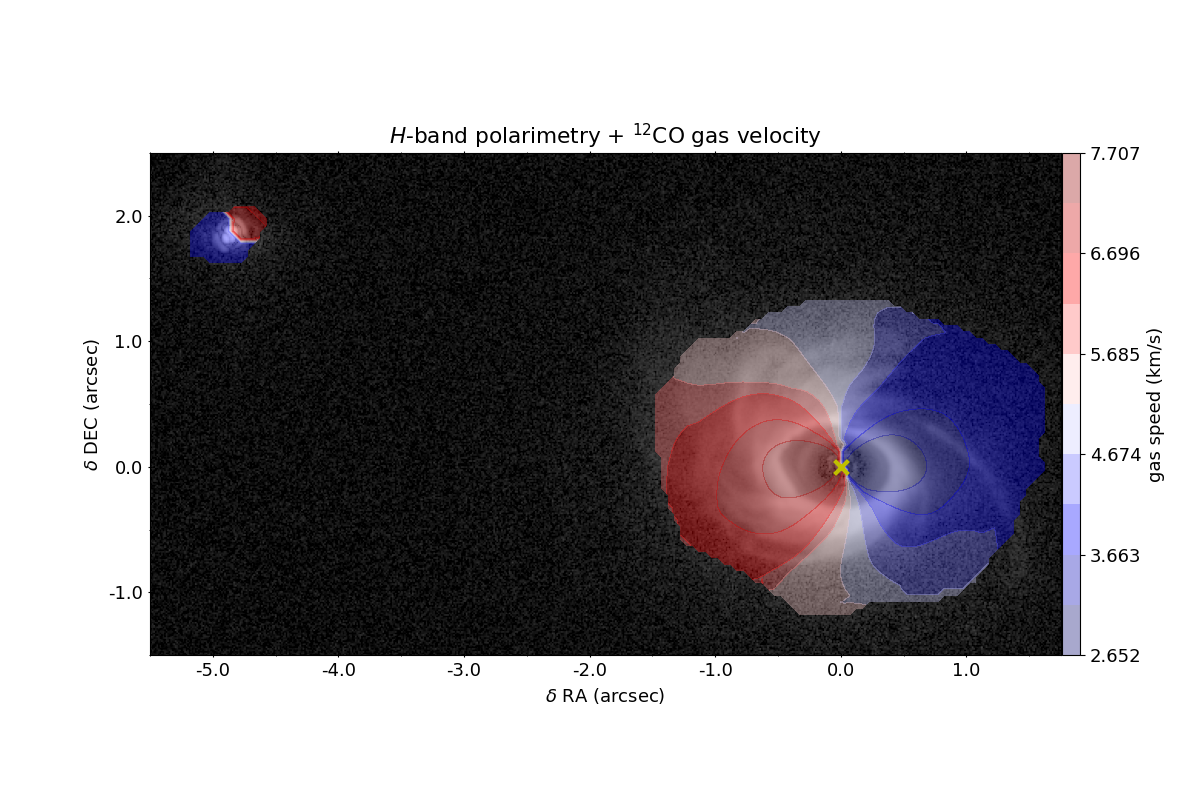}
            \caption{ ALMA data velocity map, obtained from the 12CO molecular lines, illustrating in the same FoV both HD\,34700 A and B (on the right and on the left, respectively). North is up and East is left. The yellow cross marks the A star position.
            Gas rotation is detected in the B component as well, although with a weaker SNR. The directions of rotation of the two gas disks are different, as are their PA in the sky plane. The beam is $\ang{;;0.52} \times \ang{;;0.48}$ at a PA of \ang{-61}. The gas speed values refer to the Local Standard of Rest. }
            \label{fig:COmapA+B}
        \end{figure*}

\begin{figure*}[hbt]
            \centering
            \includegraphics[trim=50 60 0 60, clip,width=\hsize]{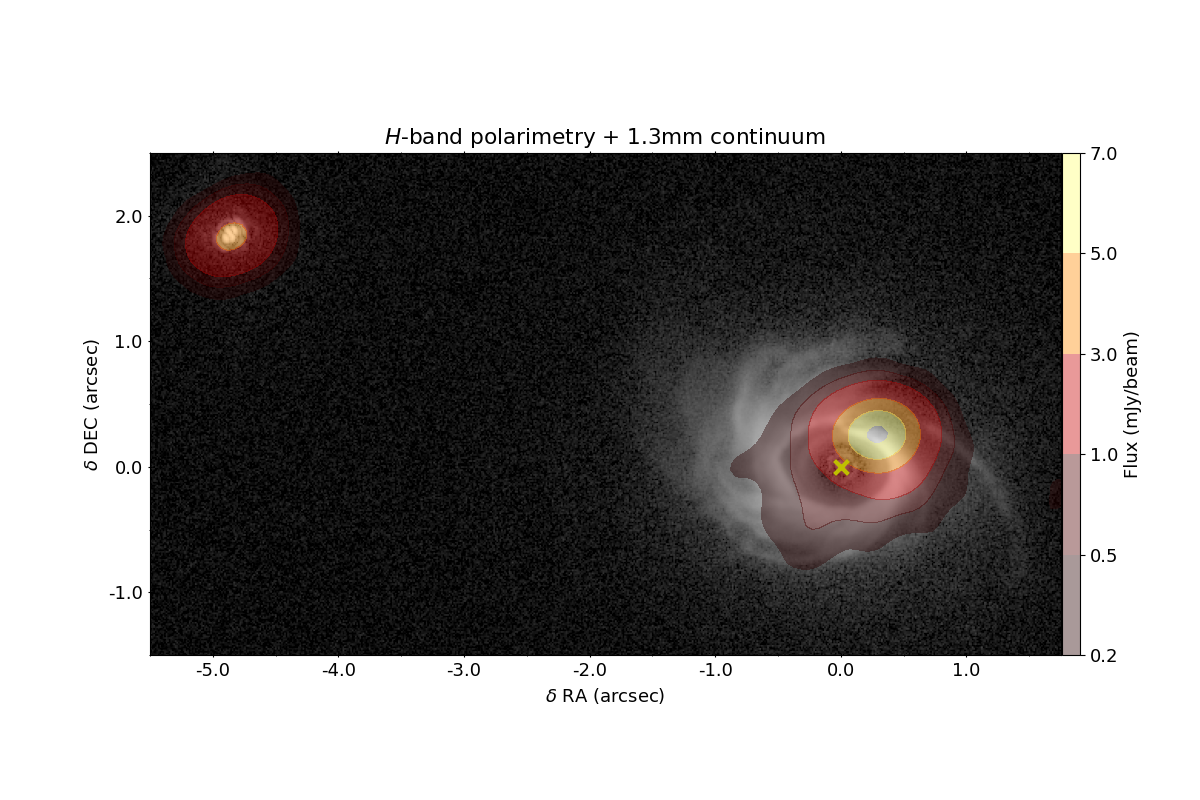}
            \caption{ ALMA 1.3\,mm continuum emission contours, including in the same FoV both HD\,34700 A and B (on the right and on the left, respectively). North is up and East is left. The yellow cross marks the A star position. The emission around A is asymmetric and off-centre, whereas around B is centred and elliptical. 
            The beam is $\ang{;;0.41} \times \ang{;;0.28}$ at a PA of \ang{-64}.  }
            \label{fig:ALMAcont_A+B}
        \end{figure*}

        Finally, in \figurename~\ref{fig:COmapA+B} and \figurename~\ref{fig:ALMAcont_A+B} we illustrate respectively the gas velocity and continuum emission obtained from the ALMA data. These images show both the A and B components of HD\,34700 in the same FoV, to highlight their differences.

        \clearpage

        \vfill 
        
        \newpage

\end{appendix}

\end{document}